\documentclass[10pt]{article}
\usepackage[tbtags]{amsmath}
\usepackage[linktocpage]{hyperref}
\hypersetup{colorlinks, linkcolor={red!60!black}, citecolor={blue!60!black}, urlcolor={blue!80!black}}
\usepackage{amssymb,amsthm,mathrsfs}
\usepackage{amsfonts}
\usepackage{graphicx}
\usepackage{stackrel}
\usepackage{subcaption}
\graphicspath{{figs/}}
\usepackage[left=2.5cm,right=2.2cm,top=0.5cm,bottom=1.3cm,includeheadfoot]{geometry}
\usepackage{indentfirst}
\usepackage{color}
\usepackage{cite}
\usepackage{mathtools}
\usepackage{tikz-cd}
\usepackage{enumitem}
\usepackage{multicol}
\usepackage{float}

\usepackage[normalem]{ulem}

\usepackage{stackengine}
\stackMath

\newcommand{\be}{\begin{equation}}
\newcommand{\ee}{\end{equation}}
\newcommand{\de}{\mbox{d}}

\newcommand{\pa}{\partial}

\newcommand{\noise}{s}

\numberwithin{equation}{section}

\newcommand{\red}{\textcolor{red}}

\renewcommand*{\thefootnote}{\fnsymbol{footnote}}

\newcommand{\doublewidetilde}[1]{\stackon[-1.9ex]{\hspace{-0.3ex}\widetilde{#1}}{\smash{\widetilde{\phantom{#1}}}}}

\begin{document}
	\begin{center}
		{\Large\bf
		Gravitational wave propagation in bigravity in the late universe\\
		}
		\vskip 5mm
		{\large
			David Brizuela$^{1,}$\footnote{e-mail address: {\tt david.brizuela@ehu.eus}},
			Marco de Cesare$^{2,3,}$\footnote{e-mail address: {\tt marco.decesare@na.infn.it}}, and Araceli Soler Oficial$^{1,}$\footnote{e-mail address: {\tt araceli.soler@ehu.eus}}}
		\vskip 3mm
		{\sl $^{1}$Department of Physics and EHU Quantum Center, University of the Basque Country UPV/EHU,\\
			Barrio Sarriena s/n, 48940 Leioa, Spain}\\\vskip 1mm
		{\sl $^2$ Scuola Superiore Meridionale, Largo San Marcellino 10, 80138 Napoli, Italy}\\\vskip 1mm
        {$^3$ INFN, Sezione di Napoli, Italy}
	\end{center}
	
\setcounter{footnote}{0}
\renewcommand*{\thefootnote}{\arabic{footnote}}
 
	\begin{abstract} 
We carry out a detailed analytical investigation of the propagation of gravitational waves in ghost-free bimetric gravity in a late-time de Sitter epoch. In this regime, the dynamical equations for the massless and massive graviton modes can be decoupled and solved exactly. We provide uniform approximations for the modes in terms of elementary functions, which are valid on all scales and for all viable mass windows. We identify different dynamical regimes for the system, depending on the propagation properties of the massive graviton, and whether the massless and massive components of the signal can be temporally resolved or not. In each regime, we compute the gravitational-wave luminosity distance as a function of redshift and study the propagation of wave packets. This allows for the derivation of a new observational bound for the ghost-free bimetric theory using the event GW170817. Further, by an explicit computation, we show that the massless and massive components of the signal retain their coherence also in the regime where they can be temporally resolved, even when couplings to incoherent matter degrees of freedom are included.
      \end{abstract}

\section{Introduction}\label{Sec:Introduction}

 Tests of gravitational wave propagation provide a powerful tool for probing new physics beyond general relativity (GR).
Modified gravity theories may introduce alterations to the standard propagation of tensor modes in a cosmological background, originating from nontrivial modifications to their dynamical equations. These may include mass terms, modified friction terms, nontrivial dispersion relations, extra polarizations, and anisotropic stress \cite{Saltas:2014dha,Nishizawa:2017nef,Ezquiaga:2018btd,Belgacem:2018lbp,LISACosmologyWorkingGroup:2019mwx,LISACosmologyWorkingGroup:2022wjo}. Such modifications leave an imprint on the gravitational wave signal, which can in principle be tested with gravitational wave observations.

A particularly relevant example is bimetric gravity \cite{Hassan:2011vm, Hassan:2011tf, Hassan:2011zd}, which extends GR by considering two independent, dynamical and nonlinearly interacting metric fields. The interactions between the two metrics are constructed in a specific way that ensures the absence of the Boulware–Deser ghost \cite{Hassan:2011tf, Hassan:2011ea}. To maintain ghost-freedom, matter fields are coupled to only one of the two metrics \cite{Yamashita:2014fga, deRham:2014naa}. The theory can also be interpreted as describing the nonlinear dynamics of two interacting spin-2 fields (gravitons), one massless and the other massive, carrying 7 dynamical degrees of freedom in total. The free parameters associated with the interaction terms in the bimetric action can be rearranged to a set of 5 physical parameters: the effective energy density of dark energy $\Omega_{\Lambda}$, the mixing angle $\theta$ between the massless and massive gravitons, the mass $m_{\rm FP}$ of the massive graviton, and two additional parameters that determine the screening mechanism on solar system scales \cite{Hogas:2021fmr,Hogas:2021lns}. Bimetric gravity admits two important limiting cases: GR is recovered in the $\theta\to0$ limit, whereas massive gravity is obtained in the limit $\theta\to\frac{\pi}{2}$. The theory is compatible with all observational tests performed so far \cite{vonStrauss:2011mq, Enander:2015kda, Platscher:2018voh, Luben:2018ekw, Hogas:2019ywm, Luben:2020xll, Caravano:2021aum, Hogas:2021lns, Hogas:2021saw, Guerrini:2023pre}, and has the potential to alleviate cosmological tensions \cite{Mortsell:2018mfj,Dwivedi:2024okk,Smirnov:2025yru,Hogas:2025ahb}.

In bimetric gravity, tensor perturbations are produced in states that are linear superpositions of massless and massive gravitons. Therefore, as they propagate, they undergo a mixing phenomenon that displays some analogies with neutrino and neutral-kaon oscillations. This was first noticed in Ref.~\cite{Hassan:2012wr} for proportional backgrounds\footnote{That is, backgrounds such that the two metrics are conformally related with constant conformal factor, $f_{ab}=C^2\,g_{ab}$. For a cosmological background, this is equivalent to having a constant ratio of the scale factors and same conformal times.}\label{Footnote:1}, and later revisited in Refs.~\cite{Max:2017flc, Max:2017kdc} for cosmological backgrounds. As discussed in Ref.~\cite{Max:2017flc}, in the regime where the gravitational wave signal component carried by the massive graviton cannot be temporally resolved from the massless component, the presence of the massive graviton induces a modulation of the gravitational wave signal, referred to as `gravitational wave oscillations'. Moreover, for gravitational wave signals with limited extent (wave packets) and sufficiently large propagation distances, the difference in group velocities between the massless and massive modes leads to a separation of the corresponding wave packets \cite{Max:2017kdc}. This has two main effects: a constant suppression factor in the amplitude of the GR-like signal carried by the massless component, and the existence of a secondary signal, interpreted as a distorted `echo', carried by the massive mode. Previous analyses have placed constraints on the theory using gravitational wave observations, excluding some regions of the $(\theta, m_{\mathrm{FP}})$-parameter space. In particular, for $m_{\mathrm{FP}} \gtrsim 10^{-23}\,\mathrm{eV}$, the mixing angle is restricted to $\theta\lesssim10^{\circ}$. Specifically, the viable regions of parameter space, which are also compatible with all the other theoretical and observational constraints obtained so far \cite{Max:2017kdc,Hogas:2021fmr, Hogas:2021lns, Hogas:2021saw,Hogas:2022owf}, include (in units of the Hubble constant $H_0$): a low-mass range $1 \lesssim m_{\rm FP}/H_0 \lesssim 10^4$; an intermediate-mass range with $10^6 \lesssim m_{\rm FP}/H_0 \lesssim 10^8$~; and a large-mass range $m_{\rm FP}/H_0 \gtrsim 10^{32}$.

In this work, we generalize the analytical studies of gravitational wave propagation in bigravity in Refs.~\cite{Max:2017flc, Max:2017kdc}, which effectively assume a Minkowski background and thus only apply to the deep-subhorizon limit and low-redshift sources, $z \ll 1$. In order to take into account more precisely cosmological effects on the propagation of the waves and extend the analysis to higher $z$, we consider the evolution of tensor modes in a de Sitter background. In this regime, modifications to standard gravitational wave propagation are governed by just two parameters: the mass and the mixing angle. Moreover, the dynamical system can be diagonalized into its mass eigenstates. We derive exact solutions for the evolution of the mass eigenstates, as well as accurate analytical approximations valid for all viable mass ranges, which reveal richer behavior that was partly missed in previous studies. We highlight the occurrence of different regimes of the dynamics depending on the value of the mass $m_{\rm FP}$ relative to the wavenumber $k$, whereby the massive graviton may either propagate (for $m_{\rm FP}\lesssim k$) or represent a localized excitation (for $m_{\rm FP}\gtrsim k$). In both regimes, we analytically compute the mode functions and the gravitational-wave luminosity distance as functions of the redshift. We also study the propagation of wave packets, which reveals some subtleties for the massive graviton in de Sitter that were not captured by previous analyses \cite{Max:2017kdc,LISACosmologyWorkingGroup:2019mwx,Ezquiaga:2021ler}. In the regime where the massive graviton is propagating, we identify two sub-
regimes, depending on whether the gravitational-wave signal components carried by the massless and massive graviton can be temporally resolved or not. Moreover, we rigorously demonstrate that both mass eigenstates originating from the same source preserve their (classical) coherence over time, also after their respective wave packets have separated---even in the presence of incoherent matter sources. This result shows that, in the context of bigravity oscillations, the term `decoherence', which is often used in the literature to describe the regime where wave packets of the mass eigenstates are nonoverlapping, and therefore can be temporally resolved \cite{Max:2017flc,Max:2017kdc}, is not accurate and shows the limitations of the analogy with neutrino oscillations.

The paper is organized as follows. In Section~\ref{Sec:Framework} we briefly review the bigravity framework and the dynamics of cosmological tensor modes in this theory. In Section~\ref{Sec:TensorModes_dS}, we study the dynamics of tensor modes in a de Sitter background. We obtain exact solutions for the mode functions of the mass eigenstates, for which we derive uniformly valid analytical approximations in terms of elementary functions. In Section~\ref{Sec:limits} we identify different regimes of physical interest based on the propagation dynamics of the massive graviton. In each regime, we compute the mode functions in terms of redshift, the gravitational-wave luminosity distance, and study the propagation of wave packets. The coherence properties of gravitational radiation in the presence of a noisy environment, represented by classical incoherent matter sources, are studied in Section~\ref{Sec:Coherence}. We conclude in Section~\ref{Sec:Discussion} by reviewing our results and discussing directions for future theoretical developments and possible applications of our results to tests of bimetric gravity using gravitational wave observations. The paper also includes two technical appendices: in Appendix~\ref{App:bessel} we detail the derivation of the asymptotic solution for the massive graviton mode, while in Appendix~\ref{App:integrals} we include details of the calculation of the correlation functions of the mass eigenstates presented in Section~\ref{Sec:Coherence}. We work with units such that the speed of light and the reduced Planck constant are both equal to one.

\section{Review of the bigravity framework}\label{Sec:Framework}

The ghost-free bimetric theory action in four dimensions is given by \cite{Hassan:2011zd}
\be\label{Eq:BimetricAction}
S_{\rm\scriptscriptstyle HR}[g_{ab},f_{ab}]=\frac{M_g^2}{2} \int \de^4 x \sqrt{-g}\, R^{(g)}+\frac{M_f^2}{2} \int \de^4 x \sqrt{-f}\, R^{(f)}-m^2 M_g^2 \int \de^4x \sqrt{-g}\,\sum_{n=0}^4 \beta_n e_n(\mathbb{S})~,
\ee
where $R^{(g)}$ and $R^{(f)}$ are the Ricci scalars of the metrics $g_{ab}$ and $f_{ab}$, respectively.  The positive constants $M_g$, $M_f$, and $m$ have physical dimensions of mass, whereas the constants $\beta_n$ are dimensionless. For later convenience, we define $\alpha\coloneqq M_f/M_g$. In particular, this ratio defines the GR-limit of the theory as $\alpha\to 0$. Finally, the $e_n(\mathbb{S})$ are the elementary symmetric polynomials of the matrix $\mathbb{S}$ \cite{Hassan:2011hr,Bernard:2015mkk}, with $\mathbb{S}$ defined in terms of the two metrics $g_{ab}$ and $f_{ab}$ as
\be
\mathbb{S}^a_{\;b}=\sqrt{g^{ac}f_{cb}}~.
\ee 
To ensure that the theory remains ghost-free in the presence of matter couplings, we assume that only $g$ interacts with matter.  For this reason, we will refer to $g$ as the physical metric. Accordingly, the full action functional has the following structure
\be\label{Eq:Action_gravity}
S[g_{ab},f_{ab},\psi]=S_{\rm\scriptscriptstyle HR}[g_{ab},f_{ab}]+S_{\rm\scriptscriptstyle m}[g_{ab},\psi]~,
\ee
where matter fields, collectively denoted by $\psi$, are minimally coupled to the metric $g$.

We consider tensor perturbations on a spatially flat Friedmann–Lemaître–Robertson–Walker cosmological background. The perturbed metrics read
\begin{subequations}\label{Eq:PerturbedFLRWmetrics}
\begin{align}
\de s_{(g)}^2&=a^2(\eta) \left[ -\de \eta^2 +\left(\gamma_{ij}+h_{ij}(\eta,x)\right)\de x^i \de x^j \right]~,\\
\de s_{(f)}^2&=b^2(\eta) \left[ -c(\eta)^2 \de \eta^2 +\left(\gamma_{ij}+\tilde{h}_{ij}(\eta,x)\right)\de x^i \de x^j \right]~,
\end{align}
\end{subequations}
where $x^{i}$ are comoving coordinates, $\eta$ denotes conformal time, and $\gamma_{ij}$ is the three-dimensional Euclidean metric. The tensor perturbations $h_{ij}$ and $\tilde{h}_{ij}$ are transverse and traceless. As in GR, linear perturbations belonging to different irreducible representations of the rotation group do not couple to one another; hence, we will not consider scalar or vector perturbations.\footnote{We note, however, that on cosmological backgrounds the scalar sector of bimetric gravity is typically plagued by a gradient instability at early times. The existence of this instability suggests a possible breakdown of linear perturbation theory rather than a physical instability, and developing a consistent framework to describe the evolution of small perturbations on cosmological backgrounds remains an open problem. For works on the instability of the scalar sector, see Refs.~\cite{Khosravi:2012rk,Berg:2012kn,Fasiello:2013woa,Comelli:2014bqa,DeFelice:2014nja,Koennig:2014ods,Kenna-Allison:2018izo,Hogas:2019ywm}. In this work, we focus on linear tensor perturbations, which are unaffected by this instability.}
We can then decompose both tensor perturbations in a polarization basis ($h_{ij}=h_+ e_{ij}^{+}+h_\times e_{ij}^{\times}$\, and $\tilde{h}_{ij}=\tilde{h}_+ e_{ij}^{+}+ \tilde{h}_\times e_{ij}^{\times}$), and the bigravity field equations imply that only components with the same polarization are coupled. Henceforth, we will drop the polarization labels $+$ and $\times$. A given polarization component is then decomposed in Fourier modes, that is, $h=\frac{1}{(2\pi)^3}\int\de\textbf{k}\, e^{i\textbf{k}\cdot\textbf{x}}h_{\textbf{k}}$\, and $\tilde{h}=\frac{1}{(2\pi)^3}\int\de\textbf{k}\, e^{i\textbf{k}\cdot\textbf{x}}\tilde{h}_{\textbf{k}}$. In momentum space, the linearized equations of motion for the Fourier components $h_{\textbf{k}}(\eta)$ and $\tilde{h}_{\textbf{k}}(\eta)$ consist of a coupled system of second-order differential equations for the two metric perturbations, 
\begin{subequations}\label{Eq:SystemTensorPerthH}
\begin{align}
&h_{\textbf{k}}^{\prime\prime}+\frac{2a^\prime}{a} h_{\textbf{k}}^{\prime}+  k^2 h_{\textbf{k}}+m^2 a^2 \lambda(y) \left(h_{\textbf{k}}-\tilde{h}_{\textbf{k}}\right) =\frac{2}{M_g^2} a^2 \pi_{\textbf{k}}~,\label{Eq:TensorPert_Harm1} \\
&\tilde{h}_{\textbf{k}}^{\prime\prime} +\left(\frac{2b^\prime}{b} -\frac{c^{\prime}}{c}\right) \tilde{h}_{\textbf{k}}^{\prime} + c^2 k^2 \tilde{h}_{\textbf{k}} +\frac{m^2}{\alpha^2} b^2 c\, y^{-4}\lambda(y) \left(\tilde{h}_{\textbf{k}}-h_{\textbf{k}}\right) =0~,\label{Eq:TensorPert_Harm2}
\end{align}
\end{subequations}
where $k=|\textbf{k}|$ is the wavenumber, a prime denotes derivative with respect to conformal time $\eta$, $y\coloneqq b/a$ is the ratio between the scale factors, and the coupling
$\lambda(y)\coloneqq \beta_1 y+\beta_2 (1+c)y^2+\beta_3 c\, y^3$ originates from the interaction terms in the bigravity action \eqref{Eq:BimetricAction}. Here, $\pi_{\textbf{k}}(\eta)$ is the Fourier transform of the anisotropic stress of matter. We note that these equations hold for any general perturbed cosmological background of the form~\eqref{Eq:PerturbedFLRWmetrics} (that is, regardless of the specific functional form of $a(\eta)$, $b(\eta)$, $c(\eta)$). For a detailed derivation of the system \eqref{Eq:SystemTensorPerthH}, we refer the reader to Refs.~\cite{Lagos:2014lca, Amendola:2015tua, Brizuela:2023uwt}. Since matter fields only couple to $h_{\textbf{k}}$, this is the only tensor mode directly relevant for gravitational wave detection.

Let us rescale $h_{\textbf{k}}$ and $\tilde{h}_{\textbf{k}}$ by their respective scale factors, in analogy with the definition of the Mukhanov-Sasaki variables in GR, we define $\mu_{\textbf{k}}\coloneqq a\, h_{\textbf{k}}$ and $\tilde{\mu}_{\textbf{k}}\coloneqq b\, \tilde{h}_{\textbf{k}}$. In terms of these variables, the system~\eqref{Eq:SystemTensorPerthH} reads
\begin{subequations}\label{Eq:MSlike-system}
    \begin{align}
&\mu_{\textbf{k}}^{\prime\prime}+\left( k^2-\frac{a^{\prime\prime}}{a}\right)\mu_{\textbf{k}}+m^2a^2 \lambda(y) \left(\mu_{\textbf{k}}-y^{-1}\tilde{\mu}_{\textbf{k}} \right)=\frac{2}{M_g^2} a^3 \pi_{\textbf{k}}~,\\
&\tilde{\mu}_{\textbf{k}}^{\prime\prime} +\left( c^2k^2-\frac{b^{\prime\prime}}{b}\right)\tilde{\mu}_{\textbf{k}}-\frac{c^{\prime}}{c}\left(\tilde{\mu}_{\textbf{k}}^{\prime}-\frac{b^{\prime}\tilde{\mu}_{\textbf{k}}}{b}\right)+\frac{m^2}{\alpha^2} c a^2 y^{-2}\lambda(y) \left(\tilde{\mu}_{\textbf{k}}-y\,\mu_{\textbf{k}} \right)=0~.
\end{align}
\end{subequations}
In Section~\ref{Sec:TensorModes_dS}, we focus on the propagation of tensor perturbations in the late universe, assuming that matter sources are inactive, i.e.,~$\pi_{\textbf{k}}=0$. The effects of matter degrees of freedom on propagation, specifically concerning their impact on the coherence of gravitational wave `echoes', are considered in Section~\ref{Sec:Coherence}.

\section{Coupled tensor modes in a de Sitter background}\label{Sec:TensorModes_dS}

In an expanding universe, viable background solutions of the bimetric field equations approach de Sitter at late times \cite{vonStrauss:2011mq}. In this limit, the relative lapse between both metrics approaches unity, $c(\eta)\to1$, and the ratio between scale factors, $y$, tends to a constant value $y_*$\cite{Comelli:2011zm}. Thus, the scale factor of the metric $g$ is
\be\label{Eq:scalefactor-dS}
a(\eta)=-\frac{1}{H\eta}~,
\ee
with constant Hubble rate $H$. The origin of time has been chosen in such a way as to have $a\to +\infty$ as $\eta\to 0^{-}$. With this convention, the range of $\eta$ is the negative half-axis. That is, we have considered that the current era of accelerated expansion continues forever. Hence, for a de Sitter background the system of equations \eqref{Eq:MSlike-system} boils down to
\begin{subequations} \label{Eq:MSlike-system-dS}
    \begin{align}
&\mu_{\textbf{k}}^{\prime\prime}+\left( k^2-\frac{2}{\eta^2}\right)\mu_{\textbf{k}}+\frac{m^2}{H^2\eta^2} \lambda(y_*) \left(\mu_{\textbf{k}}-y_*^{-1}\tilde{\mu}_{\textbf{k}} \right)=0~,\\
&\tilde{\mu}_{\textbf{k}}^{\prime\prime} +\left( k^2-\frac{2}{\eta^2}\right)\tilde{\mu}_{\textbf{k}}+\frac{m^2}{\alpha^2}\frac{\lambda(y_*)}{y_*^2H^2\eta^2} \left(\tilde{\mu}_{\textbf{k}}-y_*\,\mu_{\textbf{k}} \right)=0~.
\end{align}
\end{subequations}
It is possible to decouple the system by introducing new variables $u_{\textbf{k}}$ and $v_{\textbf{k}}$---the mass eigenstates of the system---given by
\begin{subequations}\label{Eq:TransformationToMassEigenstates}
\begin{align}
u_{\textbf{k}}&=\cos\theta\, \mu_{\textbf{k}} +\sin\theta\, \alpha \tilde{\mu}_{\textbf{k}}~,\\
v_{\textbf{k}}&=-\sin\theta\, \mu_{\textbf{k}} +\cos\theta\, \alpha \tilde{\mu}_{\textbf{k}}~,
\end{align}
\end{subequations}
with the mixing angle $\theta\in(0,\pi/2)$ defined such that\footnote{Note that this is slightly different from the definition adopted in Ref.~\cite{Luben:2020xll}, where the mixing angle is defined as $\alpha y_*$.} 
\be\label{Eq:}
\sin\theta=\frac{\alpha y_*}{\sqrt{1+\alpha^2 y_*^2}}~.
\ee
Note that the GR-limit ($\alpha\to 0$) is now encoded in terms of $\theta$ as the $\theta\to 0$ limit. We stress that the coefficients of the transformation matrix in \eqref{Eq:TransformationToMassEigenstates} do not depend on $k$; hence, the same relation \eqref{Eq:TransformationToMassEigenstates} also applies in real space. With the above transformation, we obtain that the system \eqref{Eq:MSlike-system-dS} is equivalent to
\begin{subequations}\label{Eq:MSlike-system-dS-decoupled}
\begin{align}
&u_{\textbf{k}}^{\prime\prime}+\left( k^2-\frac{2}{\eta^2}\right)u_{\textbf{k}}=0~,\label{Eq:MSlike-system-dS-decoupled1}\\
&v_{\textbf{k}}^{\prime\prime} +\left( k^2-\frac{2}{\eta^2}\right)v_{\textbf{k}}+\frac{m_{\rm FP}^2}{H^2 \eta^2} v_{\textbf{k}}=0~,\label{Eq:MSlike-system-dS-decoupled2}
\end{align}
\end{subequations}
where the Fierz-Pauli mass is defined as \cite{Luben:2020xll}
\be\label{Eq:FierzPauli-mass}
m_{\rm FP}^{2}=\frac{m^2\lambda(y_*)}{\sin^2\theta}~.
\ee
In order to ensure the absence of ghost instabilities \cite{Fasiello:2012rw,Higuchi:1986py}, we impose the Higuchi bound $m_{\rm FP}^{2}> 2 H^2$.

The system of equations \eqref{Eq:MSlike-system-dS-decoupled} can be solved exactly,
\begin{subequations}
\begin{align}
u_{\textbf{k}}(\eta) &=\frac{A_{\textbf{k}}}{-H\eta}\Big(\cos\left(-k\eta+\varphi_{\textbf{k}}\right)-k\eta\sin\left(-k\eta+\varphi_{\textbf{k}}\right)\Big)~,\label{Eq:dS-masseigenstates-sol-u}\\
v_{\textbf{k}}(\eta)&=\frac{\sqrt{-\eta}}{H}\, \Big(C_{\textbf{k}} J_{\xi}(-k \eta )+D_{\textbf{k}} Y_{\xi}(-k \eta )\Big)~,\label{Eq:dS-masseigenstates-sol-v}
\end{align}
\end{subequations}
where $A_{\textbf{k}},\,\varphi_{\textbf{k}},\,C_{\textbf{k}}$ and $D_{\textbf{k}}$ are integration constants and $J_{\xi}$ and $Y_{\xi}$ are the Bessel functions of the first and second kind, respectively. In these expressions, we introduced the shorthand notation $\xi\coloneqq\sqrt{9/4-(m_{\rm FP}/H)^2}$, and we made explicit the $k$-dependence of the integration constants. The overall $H^{-1}$ factor is just a convenient choice of normalization for the modes.
We stress that the exact solutions derived above are fully general and valid for modes on all scales and for all values of the mass. This result goes beyond previous studies, which are restricted to sub-horizon scales \cite{Max:2017flc,Max:2017kdc,Ezquiaga:2021ler}. However, for the sake of extracting concrete physical information on the behavior of the system, expression \eqref{Eq:dS-masseigenstates-sol-v} might be unwieldy. Thus, in the following, we analyze its asymptotics in a relevant regime of interest and provide an accurate approximation in terms of elementary functions. This considerably simplifies the analysis of deviations from GR and helps us better understand potential signatures of bimetric gravity in the propagation of tensor modes. The approximation derived below is also convenient in order to study the propagation of gravitational waves in real space, as analyzed in the following subsections.

As shown in Appendix~\ref{App:bessel}, using the asymptotics of the Bessel functions obtained in Ref.~\cite{dunster2025} for a purely imaginary index $\xi=i|\xi|$ with $|\xi|>1$, the solution for the $v$ mode \eqref{Eq:dS-masseigenstates-sol-v} can be approximated as 
\begin{equation}\label{Eq:dS-masseigenstates-sol-v-approx}
    v_{\textbf{k}}(\eta)= \frac{B_{\textbf{k}}\sqrt{-\eta}}{H(k^2\eta^2+|\xi|^2)^{1/4}}\cos\left(\sqrt{k^2\eta^2+|\xi|^2}+|\xi|\ln\left(\frac{-k\eta}{|\xi|+\sqrt{k^2\eta^2+|\xi|^2}}\right)+\phi_{\textbf{k}}\right)+\mathcal{O}(\Delta)~,
\end{equation}
where
\begin{equation}
   \Delta\coloneqq \frac{5\sqrt{|\xi|}-3\left(k^2\eta^2+|\xi|^2\right)^{1/4}}{24\left(k^2\eta^2+|\xi|^2\right)^{3/4}}~.
\end{equation}
The above conditions on $\xi$ correspond to the mass range $13/4<m_{\rm FP}^2/H^2$, and we will work under this assumption in the remainder of the paper. We recall that the theoretical lower bound for the mass is smaller, as it is set by the Higuchi bound, $2< m_{\rm FP}^2/H^2$. However, the mass range not covered by the above approximation, $2 < m_{\rm FP}^2/H^2\leq 13/4$, is a narrow interval and will be disregarded in the following.

Going back to the original variables, $h_{\textbf{k}}$ and $\tilde{h}_{\textbf{k}}$, we obtain that the general solution to the system \eqref{Eq:SystemTensorPerthH} can be approximated as
\begin{subequations}\label{Eq:dS-sol}
\begin{align}
    h_{\textbf{k}}(\eta)&\approx A_{\textbf{k}}\cos\theta\,\Big(\cos\left(-k\eta+\varphi_{\textbf{k}}\right)-k\eta\sin\left(-k\eta+\varphi_{\textbf{k}}\right)\Big) - B_{\textbf{k}}\sin\theta \frac{(-\eta )^{3/2}}{(k^2\eta^2+|\xi|^2)^{1/4}}\cos\left(\Phi_{\textbf{k}}(\eta)+\phi_{\textbf{k}}\right) ~,\\
    \tilde{h}_{\textbf{k}}(\eta) & \approx A_{\textbf{k}}\cos\theta\,  \Big(\cos\left(-k\eta+\varphi_{\textbf{k}}\right)-k\eta\sin\left(-k\eta+\varphi_{\textbf{k}}\right)\Big) + B_{\textbf{k}}\frac{\cos^2\theta}{\sin\theta} \frac{(-\eta )^{3/2}}{(k^2\eta^2+|\xi|^2)^{1/4}}\cos\left(\Phi_{\textbf{k}}(\eta)+\phi_{\textbf{k}}\right)~,
\end{align}
\end{subequations}
where we used $\tan{\theta}=\alpha y_*$ and introduced the shorthand notation
\be\label{Eq:phase_function}
\Phi_{\textbf{k}}(\eta)=\sqrt{k^2\eta^2+|\xi|^2}+|\xi|\ln\left(\frac{-k\eta}{|\xi|+\sqrt{k^2\eta^2+|\xi|^2}}\right)~.
\ee
These analytical approximations for the modes encapsulate all relevant physical information of gravitational wave propagation in the late universe in the theory at hand, which will be analyzed in detail in the following sections.

It is often convenient to express the evolution of tensor modes in terms of the cosmological redshift $z$. We recall that the relation between scale factor and redshift is $a(z)=a_0/(1+z)$. Combining this with Eq.~\eqref{Eq:scalefactor-dS} and choosing $a_0=1$ for the present value of the scale factor (that is, at $z=0$), we obtain the following relation between conformal time and redshift in de Sitter,
\be\label{Eq: time-redshift}
|\eta| k=\frac{k}{H}(1+z)~,
\ee
which will be useful in the following.

\subsection{Regimes of physical interest}\label{Sec:limits}

In what follows, our goal is to compute the effects of a nonvanishing graviton mass, $m_{\rm FP}\neq0$\,, on the evolution of $v_{\textbf{k}}$ depending on the value of the mass $m_{\rm FP}$ compared to the wavenumber $k$. Specifically, we are mainly interested in two different regimes, defined by two opposite hierarchies imposed on the characteristic dimensionless quantities of the system \eqref{Eq:MSlike-system-dS-decoupled}, $ m_{\rm FP}/H$ and $|\eta|k$. As will be explained below, these regimes are physically characterized as follows: (i) $ m_{\rm FP}/H\ll |\eta|k$, where the massive mode $v$ is propagating; (ii) $m_{\rm FP}/H\gg |\eta|k$, where the massive mode $v$ is nonpropagating. 
The existence of the latter regime is not surprising, given the form of the dynamical equation \eqref{Eq:MSlike-system-dS-decoupled2}, which becomes ultralocal when the momentum is much smaller than the mass (that is, the dependence on $k$ is lost in this limit). In both regimes, there exist regions of parameter space where the theory is observationally viable, specifically for small values of the mixing angle $\theta$ (see Figure~8 in Ref.~\cite{Hogas:2022owf}), although we will not make any assumptions on $\theta$ in our analysis. It is important to note that we are not assuming any specific values for the combination $|\eta| k$ and that the results derived in this section are valid on all scales. However, from the observational point of view, relevant modes must be sub-horizon (that is, $|\eta|k\gg1$) in the late-time de Sitter era.

\subsubsection{Regime 1: propagating massive graviton ($m_{\rm FP}/H\ll |\eta|k$)}\label{Sec:regime1}

This regime is defined by the condition $m_{\rm FP}/H\ll |\eta|k$\,, which translates into $|\xi|\ll|\eta|k$. Physically, it corresponds to the sub-horizon limit (i.e., $1\ll|\eta|k$), due to the Higuchi bound on the mass, $2<m_{\rm FP}^{2}/H^2 $.  Correspondingly, the physical wavelength is much shorter than the length scale $1/m_{\rm FP}$ set by the graviton mass.

In this case, the approximate solution \eqref{Eq:dS-masseigenstates-sol-v-approx} obtained earlier for the $v$ mode can be further simplified to
\begin{equation}
        v_{\textbf{k}}(\eta)\approx \frac{B_{\textbf{k}}}{H\sqrt{k}}\cos\left(k|\eta|-\frac{|\xi|^2}{2k|\eta|}+\phi_{\textbf{k}}\right)~,\quad |\xi|\ll|\eta|k~.
    \end{equation}
Hence, in terms of the redshift, the solutions for the tensor modes $h_{\textbf{k}}$ and $\tilde{h}_{\textbf{k}}$ in this regime are
\begin{subequations}\label{Eq:dS-sol-small-mass}
\begin{align}
\begin{split}
    h_{\textbf{k}}(z)&\approx A_{\textbf{k}}\cos\theta\,\left(\cos\left(\frac{k}{H}(1+z)+\varphi_{\textbf{k}}\right)+\frac{k}{H}(1+z)\sin\left(\frac{k}{H}(1+z)+\varphi_{\textbf{k}}\right)\right)\\
    &\quad - \widetilde{B}_{\textbf{k}}\sin\theta (1+z)\cos\left(\frac{k}{H}(1+z)-\frac{H|\xi|^2}{2k(1+z)}+\phi_{\textbf{k}}\right) ~,\label{Eq:dS-sol-small-mass_h}
\end{split}\\
\begin{split}
    \tilde{h}_{\textbf{k}}(z) & \approx A_{\textbf{k}}\cos\theta\,  \left(\cos\left(\frac{k}{H}(1+z)+\varphi_{\textbf{k}}\right)+\frac{k}{H}(1+z)\sin\left(\frac{k}{H}(1+z)+\varphi_{\textbf{k}}\right)\right)\\
    & \quad + \widetilde{B}_{\textbf{k}}\frac{\cos^2\theta}{\sin\theta}(1+z)\cos\left(\frac{k}{H}(1+z)-\frac{H|\xi|^2}{2k(1+z)}+\phi_{\textbf{k}}\right)~.\label{Eq:dS-sol-small-mass_htilde}
\end{split}
\end{align}
\end{subequations}
The nonstandard redshift dependence of the second term of \eqref{Eq:dS-sol-small-mass_h} is a feature of bimetric gravity in momentum space in this regime, and originates from the presence of a massive mode. We observe that this approximation is accurate for large $\xi$ as far as single monochromatic components are concerned. However, when different monochromatic components are superposed, it becomes crucial to use the unexpanded form of the argument of the trigonometric functions in order to preserve information on the relative phase shifts, as presented in Eq.~\eqref{Eq:dS-sol}. This will become apparent in the example with wave packets considered at the end of this subsection.

\vspace{0.3cm}

\underline{Gravitational-wave luminosity distance}

\vspace{0.2cm}

\begin{figure}
    \centering
    \begin{subfigure}[b]{0.48\textwidth}
    \includegraphics[width=\columnwidth]{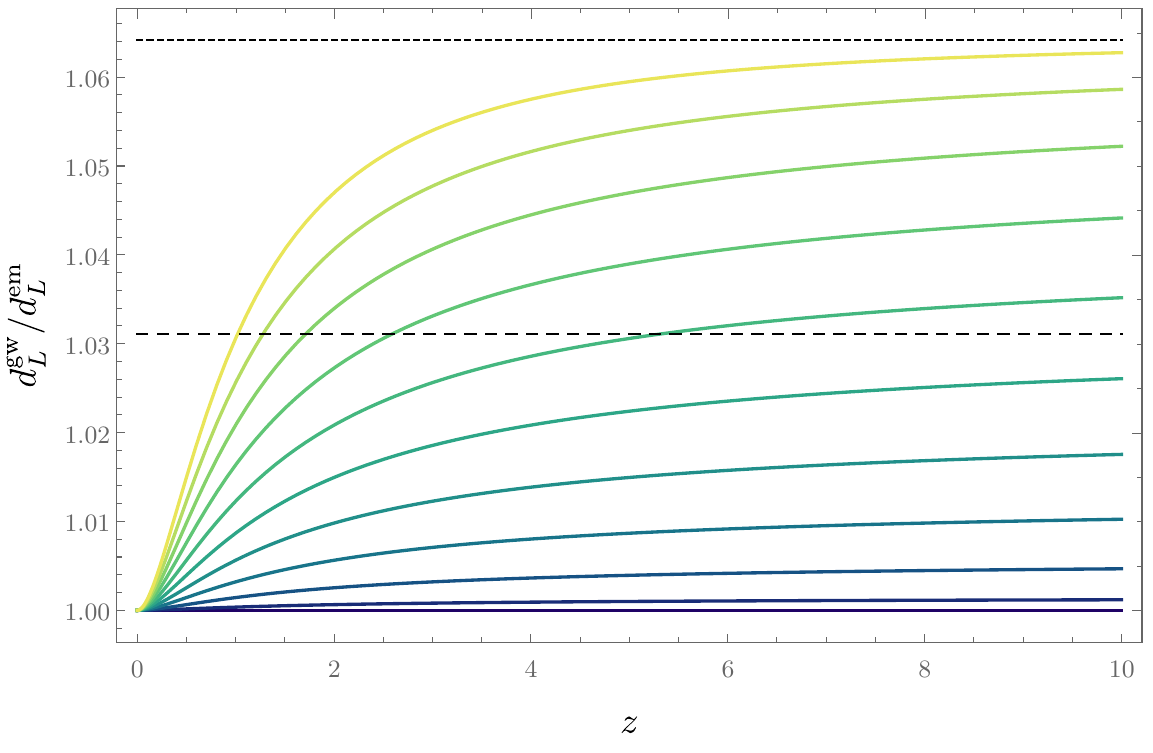}
    \caption{Below-threshold case, for masses such that $|\xi|^2\leq2\pi k/H$ and $\theta=\frac{\pi}{18}$. Lighter shades correspond to higher values of the mass. The dotted curve corresponds to the upper bound $|\cos(2\theta)|^{-1}$.
    For comparison, the dashed curve represents the `temporally resolved' regime \eqref{Eq:luminositydistance_resolvedregime}.}\label{Fig:sub1a}
    \end{subfigure}
    \hfill
    \begin{subfigure}[b]{0.48\textwidth}
    \includegraphics[width=\columnwidth]{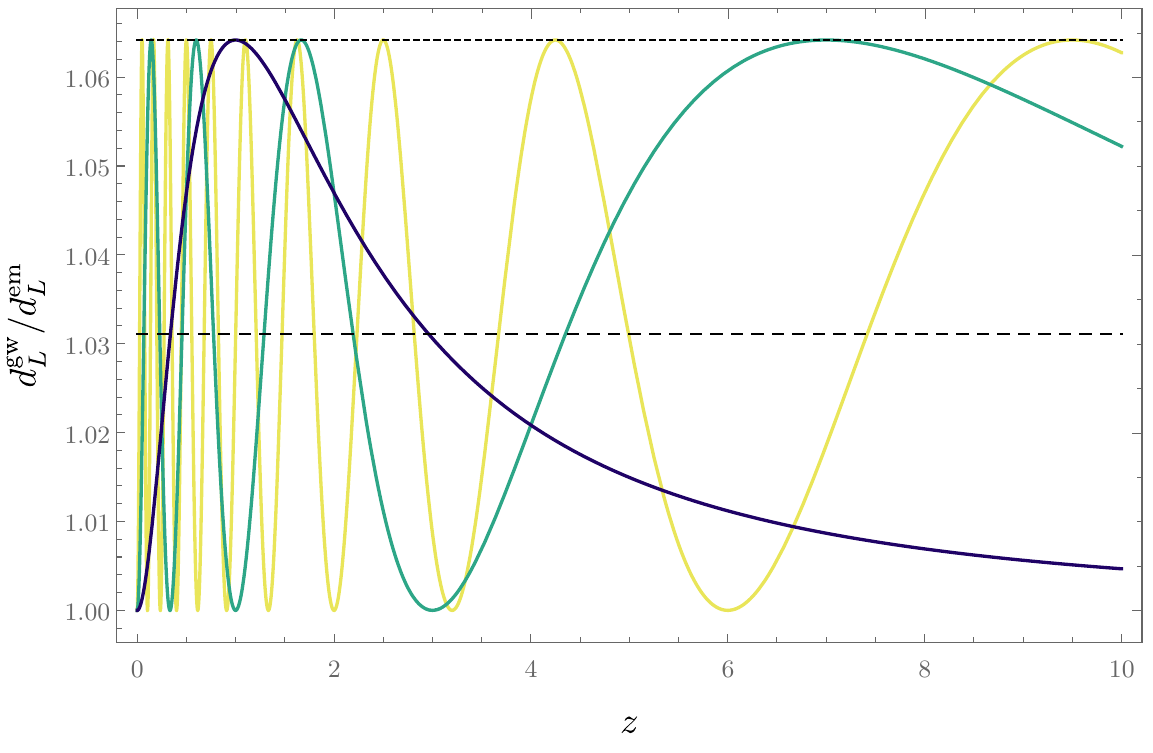}
        \caption{Above-threshold case, for masses such that $|\xi|^2>2\pi k/H$ and $\theta=\frac{\pi}{18}$. Lighter shades correspond to higher values of the mass. The dotted curve corresponds to the upper bound $|\cos(2\theta)|^{-1}$.
    For comparison, the dashed curve represents the `temporally resolved' regime \eqref{Eq:luminositydistance_resolvedregime}.}
        \label{Fig:sub1b}
    \end{subfigure}
    
    \vspace{0.5cm}  

    \begin{subfigure}[b]{0.48\textwidth}
        \includegraphics[width=\columnwidth]{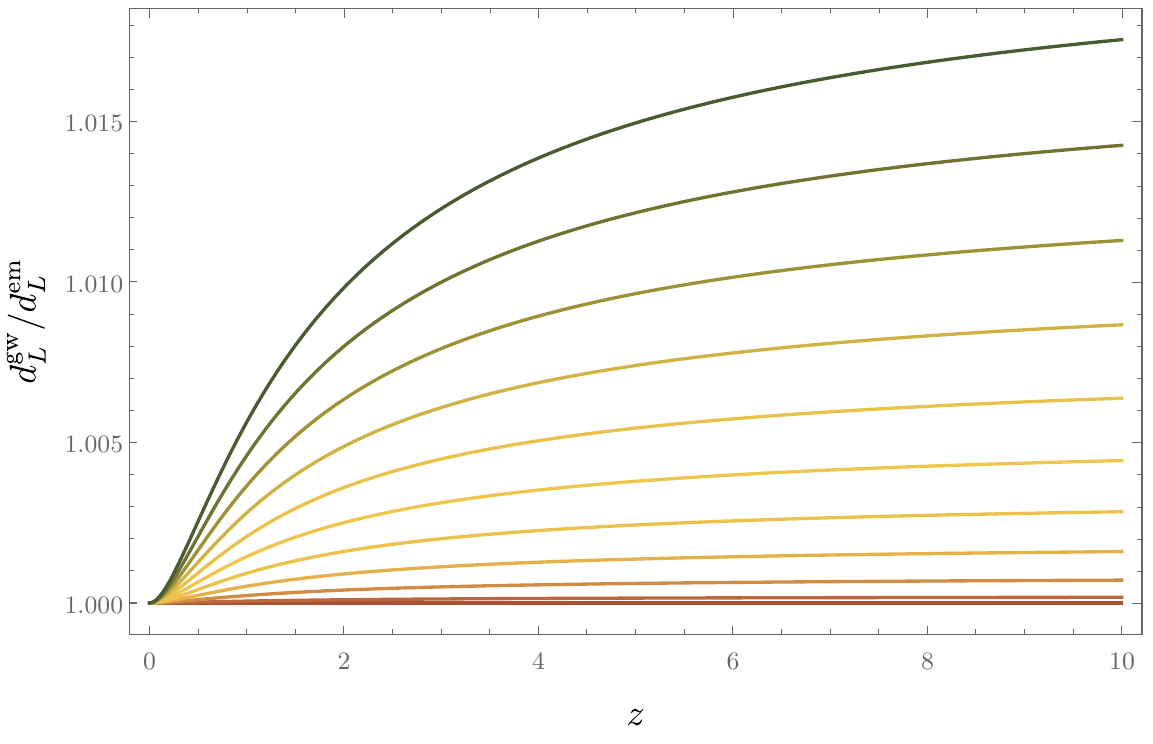}
        \caption{Below-threshold case, for masses such that $|\xi|^2\leq2\pi k/H$ and $\theta$ increasing from zero to $\frac{\pi}{18}$. Darker shades correspond to higher values of the mixing angle.}
        \label{Fig:sub1c}
    \end{subfigure}
    \hfill
    \begin{subfigure}[b]{0.48\textwidth}
        \includegraphics[width=\columnwidth]{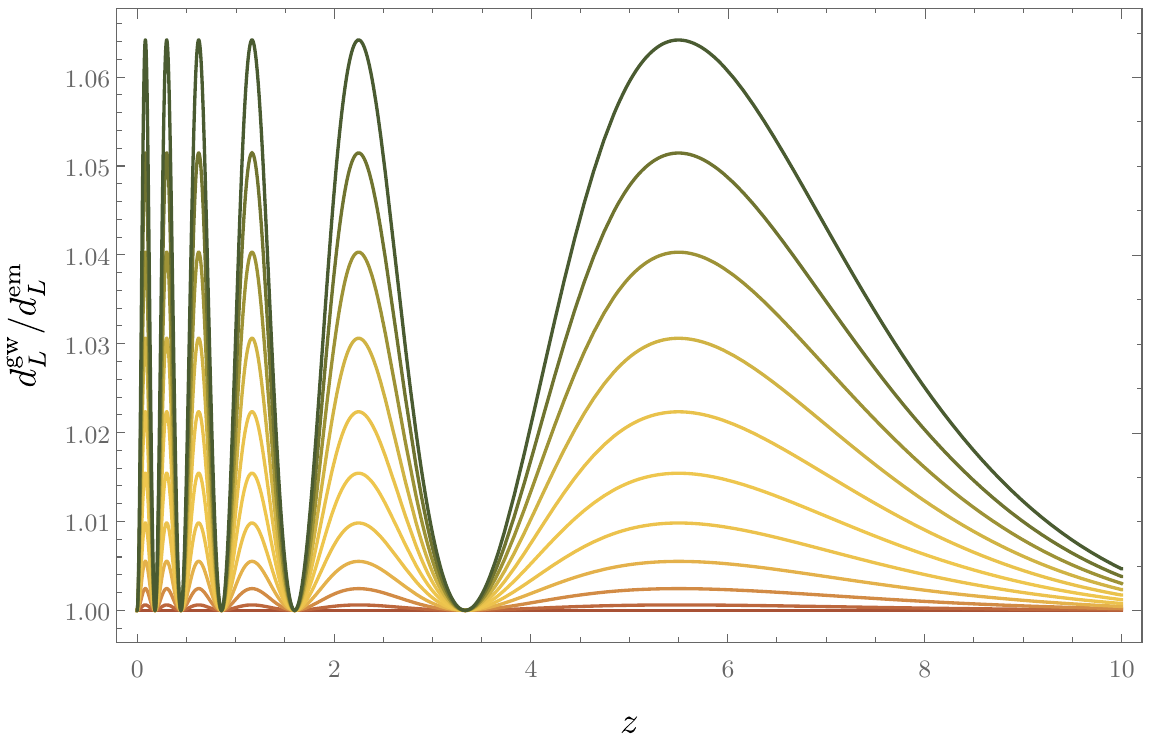}
        \caption{Above-threshold case, for masses such that $|\xi|^2>2\pi k/H$ and $\theta$ increasing from zero to $\frac{\pi}{18}$. Darker shades correspond to higher values of the mixing angle.}
        \label{Fig:sub1d}
    \end{subfigure}
    \caption{Plots of the ratio of the gravitational-wave luminosity distance to the electromagnetic luminosity distance in the `temporally unresolved' regime, as a function of the redshift $z_0$ of the emitting source given by \eqref{Eq:GW_luminositydistance}. The behavior changes significantly depending on the value of the mass. \\
    }\label{Fig:dLgw}
\end{figure}

Next, we study the effect of mixing of mass eigenstates on the gravitational-wave luminosity distance.  We compare the gravitational-wave luminosity distance in bigravity, associated with the tensor perturbation of the physical metric, $h_{\textbf{k}}(z)$, to the corresponding quantity in GR, $h^{\rm GR}_{\textbf{k}}(z)$. To make a meaningful comparison, in the two theories perturbations must be subject to the same initial conditions on the observable mode. We impose $h_{\textbf{k}}(z_0) = h^{\rm GR}_{\textbf{k}}(z_0)=f_{\textbf{k}}$ at some redshift $z_0>0$ corresponding to signal emission, and also assume as initial condition that the perturbations of the second metric are initially unexcited $\tilde{h}_{\textbf{k}}(z_0)=0$. This is expected due to the screening mechanism and the fact that only one metric couples to  matter~\cite{Max:2017kdc}.
For definiteness, we only consider positive frequency solutions, $h_{\textbf{k}}\sim e^{-ik\eta}$. Hence, using Eq.~\eqref{Eq:dS-sol-small-mass} and the above initial conditions, we obtain the following solutions in the two theories
\begin{subequations}
\begin{align}
h^{\rm GR}_{\textbf{k}}(z)&=f_{\textbf{k}}e^{i\frac{k}{H}(z-z_0)}\left(\frac{H-ik(1+z)}{H-ik(1+z_0)}\right)~,\label{Eq:GRsol_lumdis1}
\end{align}
for GR, and
\begin{align}
h_{\textbf{k}}(z)&=f_{\textbf{k}}e^{i\frac{k}{H}(z-z_0)}\left[\left(\frac{H-ik(1+z)}{H-ik(1+z_0)}\right)\cos^2\theta +\left(\frac{1+z}{1+z_0}\right)e^{\frac{i|\xi|^2H(z-z_0)}{2k(1+z)(1+z_0)}} \sin^2\theta\right]~,\label{Eq:HRsol_lumdis1}\\
\tilde{h}_{\textbf{k}}(z)&=f_{\textbf{k}}\cos^2\theta\, e^{i\frac{k}{H}(z-z_0)}\left[\left(\frac{H-ik(1+z)}{H-ik(1+z_0)}\right) -\left(\frac{1+z}{1+z_0}\right)e^{\frac{i|\xi|^2H(z-z_0)}{2k(1+z)(1+z_0)}}\right]~,\label{Eq:HRsol_lumdis1tilde}
\end{align}
\end{subequations}
for the bimetric theory.
We recall that $h_{\textbf{k}}$ is directly observable, while $\tilde{h}_{\textbf{k}}$ is not.
The gravitational-wave luminosity distance is determined by the scaling of the amplitude as a function of redshift, $|h_{\textbf{k}}|\propto 1/d_L^{\text{gw}}$ \cite{Belgacem:2017ihm,LISACosmologyWorkingGroup:2019mwx}, and in GR it coincides with the electromagnetic luminosity distance $|h^{\rm GR}_{\textbf{k}}|\propto 1/d_L^{\text{em}}$.
Computing the ratio of \eqref{Eq:GRsol_lumdis1} to \eqref{Eq:HRsol_lumdis1} in the sub-horizon limit $(k|\eta|\gg1)$, we obtain
\begin{equation}
   \frac{|h^{\rm GR}_{\textbf{k}}(z)|}{|h_{\textbf{k}}(z)|} \overset{k|\eta|\gg1}{\approx} \frac{1}{\cos^2\theta}\left[1+\tan^4\theta+2\tan^2\theta \cos\left(\frac{|\xi|^2H(z-z_0)}{2k(1+z)(1+z_0)}\right)\right]^{-1/2}~.
\end{equation}
Evaluating this quantity at present time, $z=0$, we finally obtain the gravitational-wave luminosity distance as a function of the redshift $z_0$ of the emitting source
\be\label{Eq:GW_luminositydistance}
\frac{d_L^{\text{gw}}(z_0)}{d_L^{\text{em}}(z_0)}=\frac{1}{\cos^2\theta}\left[1+\tan^4\theta+2\tan^2\theta \cos\left(\frac{|\xi|^2H z_0}{2k (1+z_0)}\right)\right]^{-1/2}~.
\ee
This expression is consistent with the result obtained in Ref.~\cite{LISACosmologyWorkingGroup:2019mwx} in the large $k$ limit using a different method (cf.~Eq.~(3.52) therein). The behavior of the luminosity distance in this regime is illustrated in Figure~\ref{Fig:dLgw}. In particular, we identify a mass threshold, corresponding to 
\begin{equation}\label{Eq:threshold}
   |\xi|^2=2\pi k/H~.
\end{equation}
For sufficiently large values of the mass, such that we can approximate $|\xi|^2\approx m_{\rm FP}^2/H^2$, the threshold \eqref{Eq:threshold} corresponds to
\begin{equation}\label{Eq:threshold_scaling}
\left(\frac{m_{\rm FP}}{10^{-22}\,{\rm eV}}\right)^2\approx 8.8\times 10^{-2}\,h \left(\frac{k}{100\,{\rm Hz}}\right)~,
\end{equation}
where $h$ is the dimensionless Hubble rate, defined according to $H= 100\, h\, {\rm km}\, {\rm s}^{-1} {\rm Mpc}^{-1}$. The behavior of the luminosity distance varies significantly depending on whether the mass lies above or below this threshold. Specifically, on the one hand, for values of the mass such that $|\xi|^2\leq2\pi k/H$, the ratio of luminosity distances \eqref{Eq:GW_luminositydistance} is a monotonically increasing function of the redshift $z_0$  and, for large redshift values, it approaches the limit 
\begin{equation}
  \lim_{z_0\to+\infty}  \frac{d_L^{\text{gw}}(z_0)}{d_L^{\text{em}}(z_0)}=\frac{1}{\cos^2\theta} \left[1+\tan^4\theta+2\tan^2\theta \cos\left(\frac{|\xi|^2H}{2k}\right)\right]^{-1/2}~.
\end{equation} 
On the other hand, for $|\xi|^2>2\pi k/H$, \eqref{Eq:GW_luminositydistance} transitions to an oscillatory behavior, subject to the bounds
\begin{equation}\label{Eq:luminositydistance-bounds}
    1\leq d_L^{\text{gw}}(z_0)/d_L^{\text{em}}(z_0)\leq |\cos(2\theta)|^{-1}~.
\end{equation}
 For $\theta\neq\frac{\pi}{4}$, the total number of peaks and troughs is $\left\lceil \frac{H|\xi|^2}{2\pi k} -1\right\rceil$. Interestingly, the upper bound is divergent when the mixing angle is $\theta=\frac{\pi}{4}$ (maximal mixing). Physically, such zeroes stem from the fact that, in the regime at hand and for $\theta=\pi/4$\,, Eq.~\eqref{Eq:HRsol_lumdis1} represents a superposition of two equal-amplitude waves with a small $z_0$-dependent beat frequency. While the GR amplitude $|h^{\rm GR}_{\textbf{k}}(z=0)|$ is always nonzero, the amplitude of the bigravity signal $|h_{\textbf{k}}(z=0)|$ vanishes at some $z_0>0$ where the massive and massless signal components interfere destructively. Hence, this divergence simply reflects complete destructive interference in the maximally mixed case.
 
 We remark that for sub-horizon modes, $|\xi|\ll|\eta|k$, the threshold \eqref{Eq:threshold} lies well within the accessible range of $|\xi|$ and is therefore physically relevant.
 The existence of two different regimes can be understood intuitively as follows. Below the threshold \eqref{Eq:threshold}, the relative phase between the massless and massive components is a monotonically increasing function of redshift and strictly less than $\pi$, even for an infinitely distant source (in which case it tends asymptotically to a constant). Hence, even though destructive interference between the two components implies an increase of $d_L^{\text{gw}}(z_0)/d_L^{\text{em}}(z_0)$ with redshift, the behavior is monotonic and bounded from above. On the other hand, above the threshold \eqref{Eq:threshold}, the relative phase exceeds $\pi$, leading to a succession of phases where $d_L^{\text{gw}}(z_0)/d_L^{\text{em}}(z_0)$ is monotonically increasing or decreasing. We note that the existence of the threshold \eqref{Eq:threshold} itself is a peculiarity of the nonlinear dependence of the massive mode on the redshift.

Finally, we observe that the luminosity distance \eqref{Eq:GW_luminositydistance} cannot be modelled using the phenomenological parametrization proposed in Ref.~\cite{Belgacem:2018lbp}, which assumes
\be\label{Eq:GW_luminositydistance_pheno}
\left(\frac{d_L^{\text{gw}}(z_0)}{d_L^{\text{em}}(z_0)}\right)^{\rm pheno}=\Xi_0+\frac{1-\Xi_0}{(1+z_0)^n}~,
\ee
with free parameters $\Xi_0$, $n$. The two formulas \eqref{Eq:GW_luminositydistance} and \eqref{Eq:GW_luminositydistance_pheno} are clearly incompatible in the above-threshold regime, due to the oscillatory nature of the bigravity luminosity distance. They are also incompatible in the below-threshold regime: even though at high redshift both expressions asymptote to a constant value, and both are monotonically increasing for $n>0$ and $\Xi_0>1$, their functional forms are clearly different. Moreover, they predict a different behavior of the luminosity distance at low redshift, where we have the following asymptotics
\begin{subequations}
\begin{align}
&\frac{d_L^{\text{gw}}(z_0)}{d_L^{\text{em}}(z_0)}\approx 1+\frac{H^2|\xi|^4}{32k^2}\sin^2(2\theta)z_0^2+{\cal O}(z_0)^3~,~\label{Eq:small_redshift_dL}\\
&\left(\frac{d_L^{\text{gw}}(z_0)}{d_L^{\text{em}}(z_0)}\right)^{\rm pheno}\approx 1+ n(\Xi_0-1)z_0+{\cal O}(z_0)^2~,
\end{align}
\end{subequations}
which are incompatible to leading order in $z_0$. Moreover, we note that $d_L^{\text{gw}}$ in bigravity depends on the wave-number $k$, whereas the phenomenological model \eqref{Eq:GW_luminositydistance_pheno} assumes that $\Xi_0$, $n$ do not dependent on $k$. We conclude that observational constraints based on the phenomenological parametrization \eqref{Eq:GW_luminositydistance_pheno} cannot be translated into constraints on the parameter space of bigravity.

The result \eqref{Eq:GW_luminositydistance} is valid in the regime where the massless and massive components of the signal cannot be temporally resolved at the detector (`temporally unresolved' regime), since the observed signal is described by the full expression in Eq.~\eqref{Eq:HRsol_lumdis1}. In turn, in the regime where the massless and massive components can be resolved at the detector (`temporally resolved' regime), the luminosity distance is determined only by the massless component, represented by the first term in the solution~\eqref{Eq:HRsol_lumdis1}. (As explained later in this section, the massive component represents a delayed `echo' of the original signal.) The amplitude of the massless component is suppressed by a factor $\cos^2\theta$ compared to the corresponding solution in GR. Hence, the luminosity distance in the `temporally resolved' regime reads
\be\label{Eq:luminositydistance_resolvedregime}
\frac{d_L^{\text{gw}}(z_0)}{d_L^{\text{em}}(z_0)}=\frac{1}{\cos^2\theta}~.
\ee

\begin{figure}
 \centering
\begin{subfigure}[b]{0.48\textwidth}
\includegraphics[width=\columnwidth]{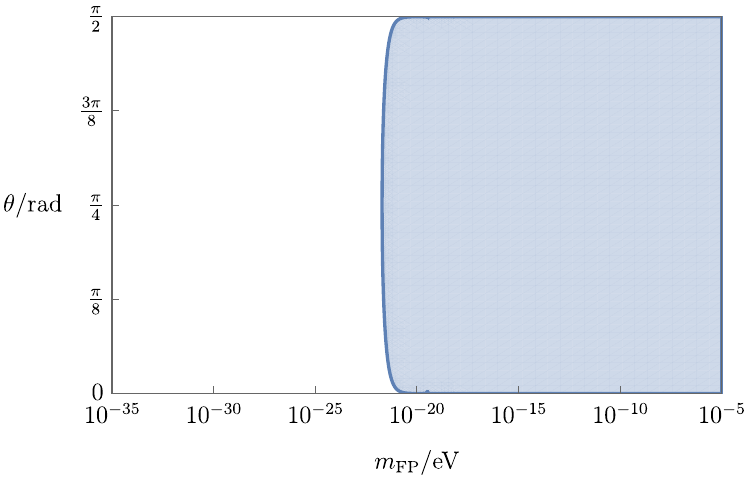}
\end{subfigure}
\begin{subfigure}[b]{0.48\textwidth}
\includegraphics[width=\columnwidth]{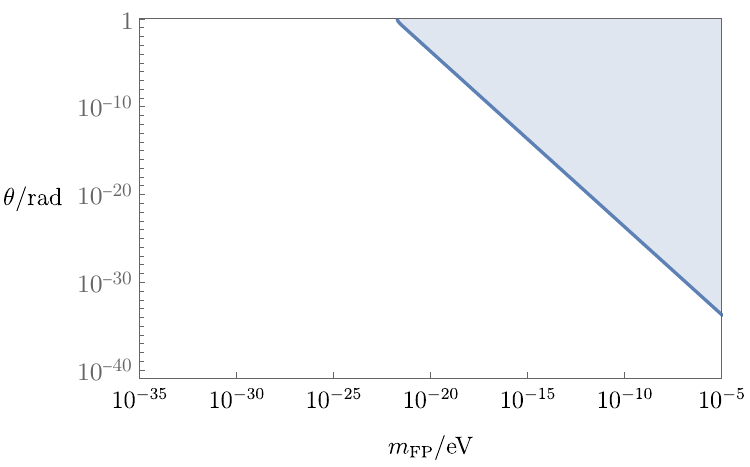}
\end{subfigure}
        \caption{The shaded area represents the region of parameter space $(\theta, m_{\mathrm{FP}})$ ruled out by the constraint \eqref{Eq:constraint-theta-m-plane}, for $h=0.7$\,. The left panel shows the excluded region for $\theta\in(0,\pi/2)$ in a linear scale, while the right panel shows a close-up for $\theta\in(10^{-40},1)$ in a logarithmic scale to better illustrate the behavior of the contour for small angles.}\label{Fig:ExclusionPlot}
\end{figure}

Multimessenger events, based on the combined observation of gravitational-wave and electromagnetic signals emitted from the same astrophysical source, can be used to constrain the bigravity parameter space $(\theta,m_{\rm FP})$. For illustrative purposes, we focus on the low-redshift event GW170817 \cite{LIGOScientific:2017vwq} observed by LIGO-Virgo detectors and its electromagnetic counterpart. We start by observing that, for sufficiently large values of $m_{\rm FP}$, we can approximate $|\xi|\approx m_{\rm FP}/H$ in Eq.~\eqref{Eq:small_redshift_dL}, which can then be recast as
\be\label{Eq:MassSinThetaFormula_smallz}
m^2_{\rm FP}\sin(2\theta)=\frac{4\sqrt{2}\,H k}{z_0}\sqrt{\frac{d_L^{\text{gw}}}{d_L^{\text{em}}}-1}~.
\ee
For GW170817, we have $d_L^{\text{gw}}=43.8^{+2.9}_{-6.9}\,{\rm Mpc}$, $d_L^{\text{em}}=40.7\pm 1.4\,{\rm (stat)}\pm 1.9\,{\rm (sys)}\,{\rm Mpc}$, $z_0\simeq0.0098$ \cite{Belgacem:2018lbp,Finke:2021aom,LIGOScientific:2017adf,Cantiello:2018ffy}. We consider the minimal frequency of the signal, which corresponds to $k=2\pi\times 24\, {\rm Hz}$ \cite{LIGOScientific:2017vwq}. Using Eq.~\eqref{Eq:MassSinThetaFormula_smallz}  with these experimental data, and combining the experimental errors in quadrature,\footnote{A similar approach was followed in Refs.~\cite{Belgacem:2018lbp,Finke:2021aom} to derive constraints on the phenomenological model \eqref{Eq:GW_luminositydistance_pheno}.} we obtain the constraint $(m_{\rm FP}^2/H)\sin(2\theta)< 2.6\times 10^{-11}\,{\rm eV}$. This can be re-expressed in terms of the reduced Hubble rate $h$ as
\begin{equation}\label{Eq:constraint-theta-m-plane}
    \left(\frac{m_{\rm FP}}{10^{-22}\,{\rm eV}}\right)^2\sin(2\theta)< 5.5\, h ~.
\end{equation}
The constraint \eqref{Eq:constraint-theta-m-plane} still allows for significant deviations of $d_L^{\text{gw}}/d_L^{\text{em}}$ from GR at large redshift. Importantly, it is also much milder compared to bounds on $m_{\rm FP}$ in massive gravity, due to the extra freedom represented by the mixing angle. An exclusion plot illustrating the constraint \eqref{Eq:constraint-theta-m-plane} in the $(\theta, m_{\mathrm{FP}})$ plane is shown in Figure~\ref{Fig:ExclusionPlot}. We recall that this bound has been derived under the assumption that the source distance is such that the massless and massive signal components cannot be temporally resolved at the detector. Further constraints based on the non-observation of waveform distortions in the `temporally unresolved' regime, as well as amplitude suppression and echoes in the `temporally resolved' regime, have been analyzed in Refs.~\cite{Max:2017flc,Max:2017kdc,Hogas:2022owf} (where such regimes are referred to as `coherent' and `decoherent', respectively). These are complementary to the bound \eqref{Eq:constraint-theta-m-plane}.

\vspace{0.3cm}

\underline{Propagation of wave packets}

\vspace{0.2cm}

To illustrate further phenomenological features of gravitational wave propagation in this regime in real space, let us consider a simple one-dimensional example subject to the initial conditions specified above, and assume that both mass eigenstates are propagating along the positive $x$ direction. The analysis of propagation in real space will also clarify the transition from the regime where the massless and massive components of the signal cannot be resolved, to the regime where they can be resolved.
We assume the following profile for $f_{\textbf{k}}$, peaked at wavenumber $k_0>0$
\be\label{Eq:initialconditionAk}
f_{\textbf{k}}=N\left(1-\frac{i k}{H}(1+z_0) \right)e^{- \frac{\sigma^2}{2}(k-k_0)^2}e^{-ik x_0}~.
\ee
Here $N$ is a normalization constant, $\sigma$ represents the width of the wave packet in real space, and $x_0$ is the position of the emitting source.
For the sake of this example, it is more convenient to study the evolution in terms of conformal time $\eta$. 
After transforming the solutions \eqref{Eq:HRsol_lumdis1}--\eqref{Eq:HRsol_lumdis1tilde}
in momentum space to real space and computing the real part of the Fourier integrals, $h(\eta,x)=\Re\left[\frac{1}{2\pi}\int_{-\infty}^{+\infty}\de k\;h_{\textbf{k}}(\eta) e^{ik x}\right]$ and $\tilde{h}(\eta,x)=\Re\left[\frac{1}{2\pi}\int_{-\infty}^{+\infty}\de k\;\tilde{h}_{\textbf{k}}(\eta) e^{ik x}\right]$, the solutions
are expressed as the linear combination,
\begin{subequations}\label{Eq:wave-packet-sol_smallmass}
\begin{align}
h(\eta,x)&=\cos^2\theta\,h_{1}(\eta,x)+\sin^2\theta\, h_{2}(\eta,x)~,\label{Eq:wave-packet-sol_smallmass_1}\\
\tilde{h}(\eta,x)&=\cos^2\theta\left(h_{1}(\eta,x)-h_{2}(\eta,x)\right)~.\label{Eq:wave-packet-sol_smallmass_2}
\end{align}
\end{subequations}
in terms of a massless $h_1$ and massive $h_2$ waves.\footnote{Note that $h_1$ and $h_2$ are simply a rescaling of $u$ and $v$, that is,
 $u=a\cos\theta\,h_1$ and $v=-a\sin\theta\,h_2$.}

The Fourier integral that defines the massless mode $h_1$ in Eq.~\eqref{Eq:wave-packet-sol_smallmass} can be computed exactly and reads as
\be\label{Eq:wave-packet-sol_smallmass-h1}
h_{1}(\eta,x)=\frac{N\, e^{-\frac{U^2}{2\sigma^2 }}}{\sqrt{2\pi}\sigma}\Big[ \left(1+\frac{\eta U}{\sigma^2}\right)\cos\left(k_0 U\right)+k_0\eta\sin\left(k_0 U\right)\Big]~,
\ee
where $U=\eta-x-\eta_0+x_0$ is a retarded time coordinate.
In order to calculate the corresponding integral for the massive mode $h_2$ analytically, we assume that the parameter $\sigma$ that appears in Eq.~\eqref{Eq:initialconditionAk} is large (specifically, $\sigma\gg1/k_0$, which ensures that $f_{\textbf{k}}$ is sharply peaked in momentum space at $k_0$). Under this assumption, we can make a standard approximation and linearize the argument of the mode functions, which enables us to compute the Fourier integral analytically. There is a subtle though important aspect that must be taken into account in the calculation: to obtain the correct behavior of the wave packet, particularly concerning its group velocity, we must use the full form of the phase of the mode functions as given in Eq.~\eqref{Eq:phase_function}.
We have
\be\label{Eq:h2solution_smallmass}
\begin{split}
&h_2(\eta,x)=\Re\left[\int\limits_{-\infty}^{+\infty}\frac{\de k}{2\pi}\left(\frac{\eta}{\eta_0}\right)\left(1+i k \eta_0 \right)e^{i\big(\Phi_{\textbf{k}}(\eta)-\Phi_{\textbf{k}}(\eta_0)+k(x-x_0)\big)}N e^{- \frac{\sigma^2}{2}(k-k_0)^2}\right]\\
&\approx  N\left(\frac{\eta}{\eta_0}\right)\Re\left[e^{i\big(\Phi_{k_0}(\eta)-\Phi_{k_0}(\eta_0)+k_0(x-x_0)\big)} \int\limits_{-\infty}^{+\infty}\frac{\de k}{2\pi} \left(1+i k \eta_0 \right)
e^{i(k-k_0)\left(\sqrt{\eta^2+\frac{|\xi|^2}{k_0^2}}-\sqrt{\eta_0^2+\frac{|\xi|^2}{k_0^2}}+x-x_0\right)} e^{- \frac{\sigma^2}{2}(k-k_0)^2}\right]\\
&=\frac{N\, e^{-\frac{W^2}{2\sigma^2 }}}{\sqrt{2\pi}\sigma}\left[ \left(\left(\frac{\eta}{\eta_0}\right)-\frac{\eta W}{\sigma^2}\right) \cos(k_0\widetilde{W}) -k_0\eta\sin(k_0\widetilde{W}) \right]~,
\end{split}
\ee
where we defined
\begin{subequations}
\begin{align}
W&=\sqrt{\eta^2+\frac{|\xi|^2}{k_0^2}}-\sqrt{\eta_0^2+\frac{|\xi|^2}{k_0^2}}+x-x_0~,\label{Eq:Wdefinition}\\
\widetilde{W}&=\sqrt{\eta^2+\frac{|\xi|^2}{k_0^2}}-\sqrt{\eta_0^2+\frac{|\xi|^2}{k_0^2}}+x-x_0+\frac{|\xi|}{k_0}\ln\left(\frac{\eta}{\eta_0}\frac{|\xi|+\sqrt{\eta_0^2+|\xi|^2}}{|\xi|+\sqrt{\eta^2+|\xi|^2}}\right).\label{Eq:Wtildedefinition}
\end{align}    
\end{subequations}

\begin{figure}[tbp]
    \centering
     \begin{subfigure}[b]{0.31\textwidth}
    \includegraphics[width=\columnwidth]{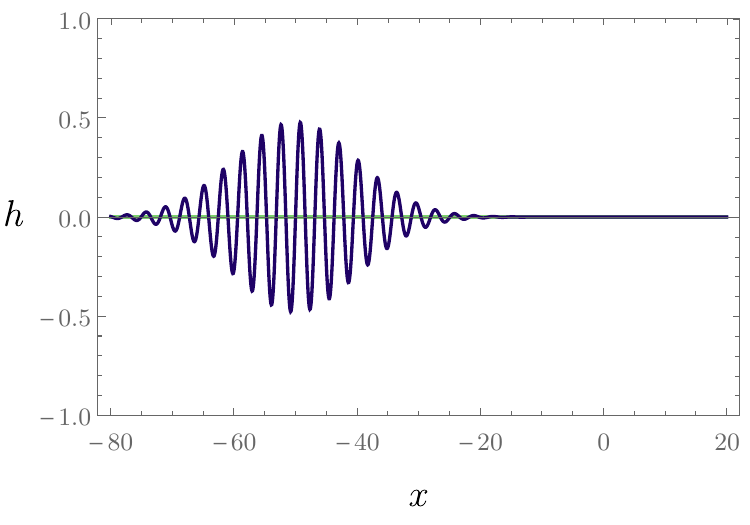}
    \caption{Initial state at $\eta=\eta_0$.}\label{Fig:sub3a}
    \end{subfigure}
      \hfill
 \begin{subfigure}[b]{0.31\textwidth}
    \includegraphics[width=\columnwidth]{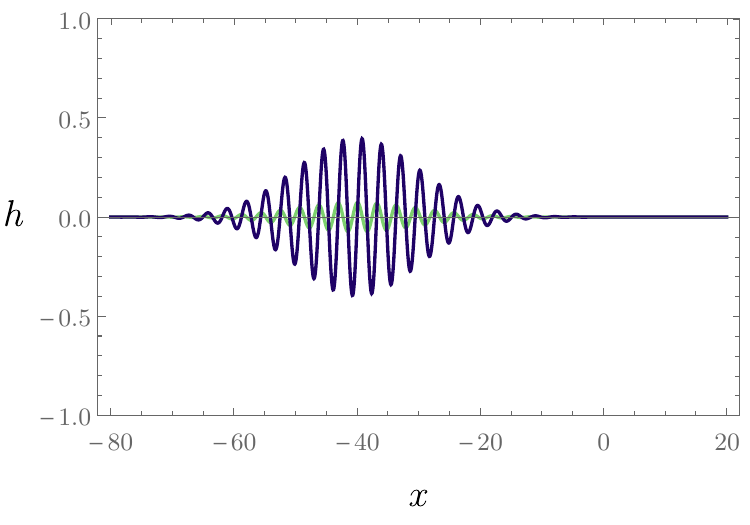}
    \caption{Evolved state at $\eta=\eta_0+\sigma$.}\label{Fig:sub3b}
    \end{subfigure}
      \hfill
    \begin{subfigure}[b]{0.31\textwidth}
    \includegraphics[width=\columnwidth]{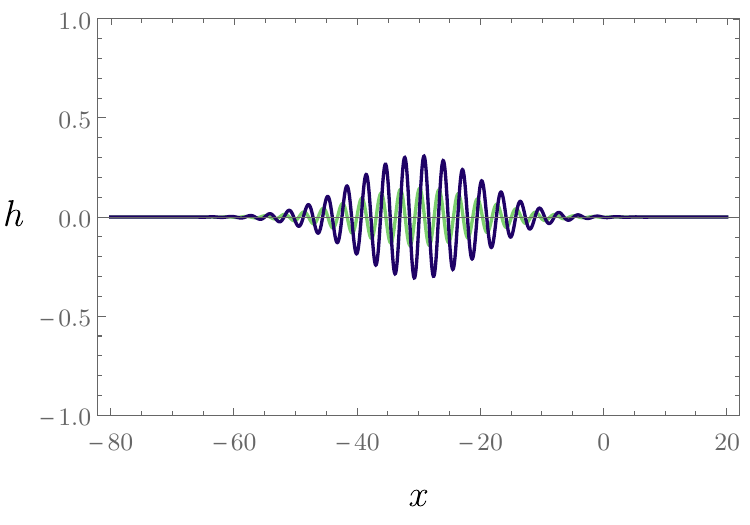}
    \caption{Evolved state at $\eta=\eta_0+2\sigma$.}\label{Fig:sub3c}
    \end{subfigure}
    \vspace{0.5cm} 
    \begin{subfigure}[b]{0.31\textwidth}
    \includegraphics[width=\columnwidth]{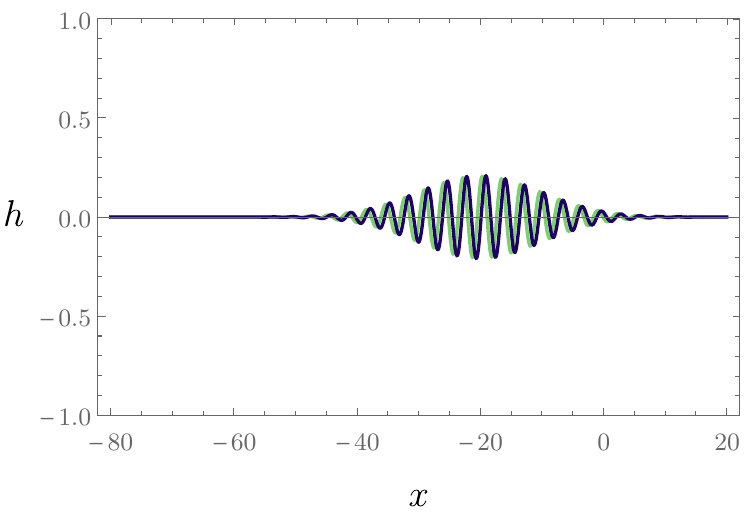}
        \caption{Evolved state at $\eta=\eta_0+3\sigma$.}
        \label{Fig:sub3d}
    \end{subfigure}
    \hfill
    \begin{subfigure}[b]{0.31\textwidth}
        \includegraphics[width=\columnwidth]{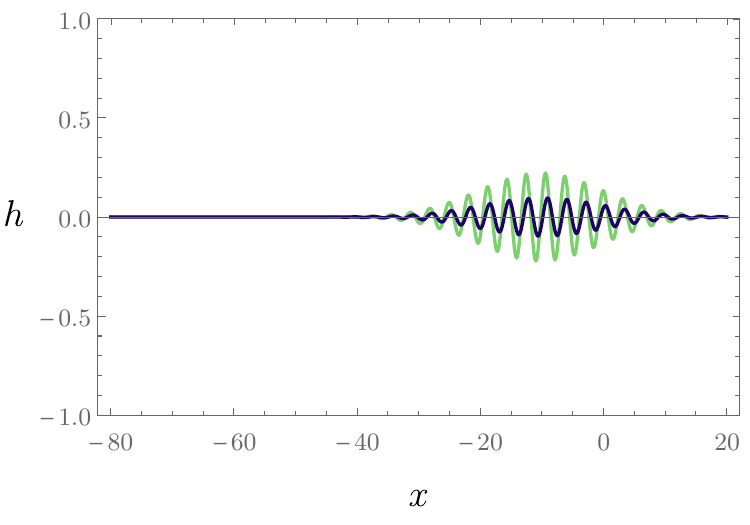}
        \caption{Evolved state at $\eta=\eta_0+4\sigma$.}
        \label{Fig:sub3e}
    \end{subfigure}
    \hfill
    \begin{subfigure}[b]{0.31\textwidth}
        \includegraphics[width=\columnwidth]{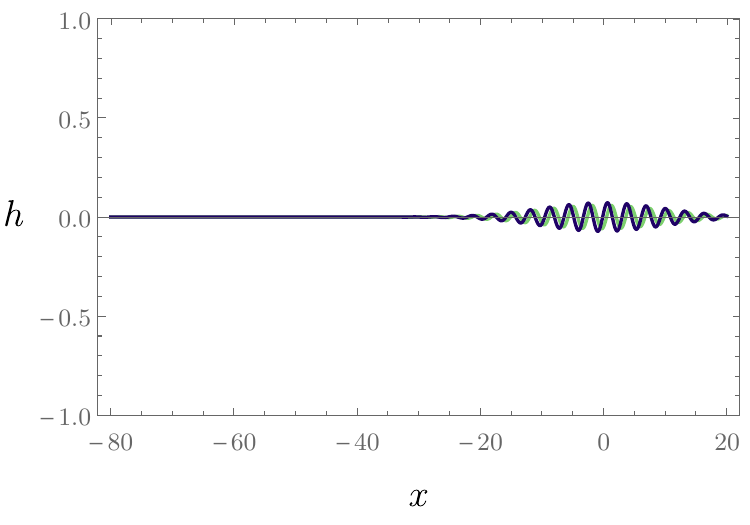}
        \caption{Evolved state at $\eta=\eta_0+5\sigma$.}
        \label{Fig:sub3f}
    \end{subfigure}
    \vspace{-0.5cm}
\caption{The figure depicts the evolution of wave-packet solutions for the $h$ (blue curve) and $\tilde{h}$ (green curve) modes in the `propagating regime' in the `temporally unresolved' case, for initial data such that $\tilde{h}(\eta_0,x)=0$ and only right-moving modes present, corresponding to the analytical solutions \eqref{Eq:wave-packet-sol_smallmass}, with parameters $|\xi|=10$, $\theta=\pi/6$, $H=10^{-1}$, $k_0=20 H$, $\sigma=10$, $x_0=-50$, $\eta_0=-60$,   $N=H$. The plots above are equally spaced in conformal time, starting at $\eta=\eta_0$ and increasing in time steps $\Delta\eta=\sigma$. The dominant effect that characterizes the propagation in real space is an oscillation phenomenon, which stems from the nontrivial relation (mixing) between $h$ and $\tilde{h}$ with the mass eigenstates.}\label{Fig:wavepackets_smallmass}
\end{figure}

\begin{figure}[tbp]
    \centering
     \begin{subfigure}[b]{0.31\textwidth}
    \includegraphics[width=\columnwidth]{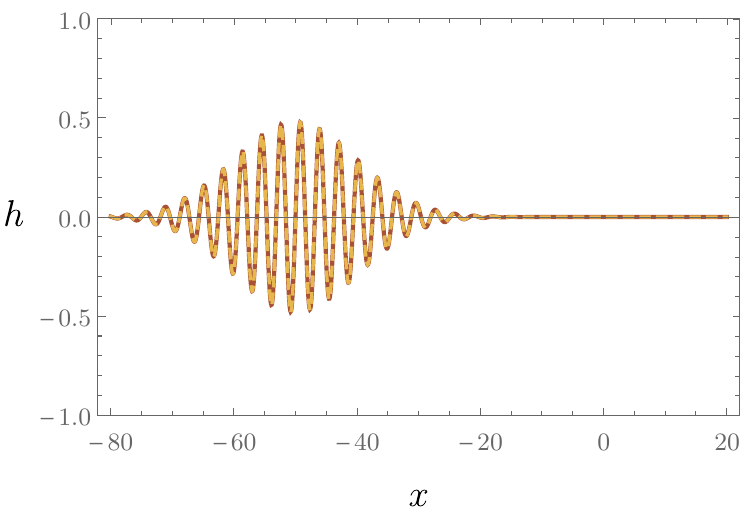}
    \caption{Initial state at $\eta=\eta_0$.}\label{Fig:sub4a}
    \end{subfigure}
      \hfill
 \begin{subfigure}[b]{0.31\textwidth}
    \includegraphics[width=\columnwidth]{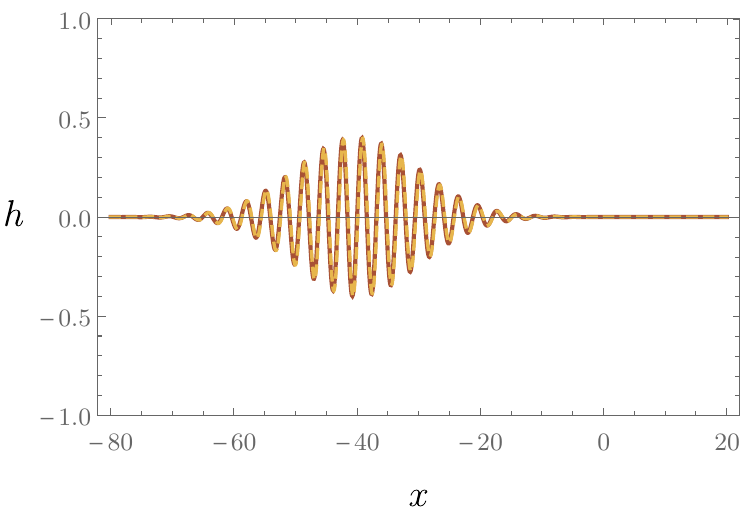}
    \caption{Evolved state at $\eta=\eta_0+\sigma$.}\label{Fig:sub4b}
    \end{subfigure}
      \hfill
    \begin{subfigure}[b]{0.31\textwidth}
    \includegraphics[width=\columnwidth]{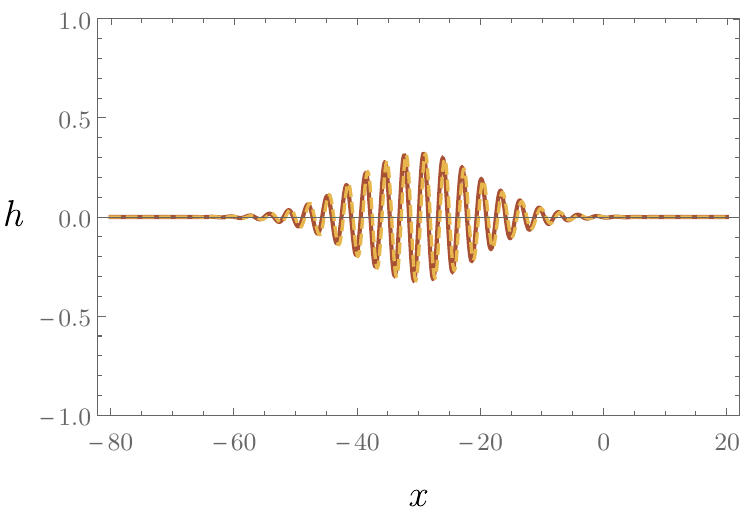}
    \caption{Evolved state at $\eta=\eta_0+2\sigma$.}\label{Fig:sub4c}
    \end{subfigure}
    \vspace{0.5cm} 
    \begin{subfigure}[b]{0.31\textwidth}
    \includegraphics[width=\columnwidth]{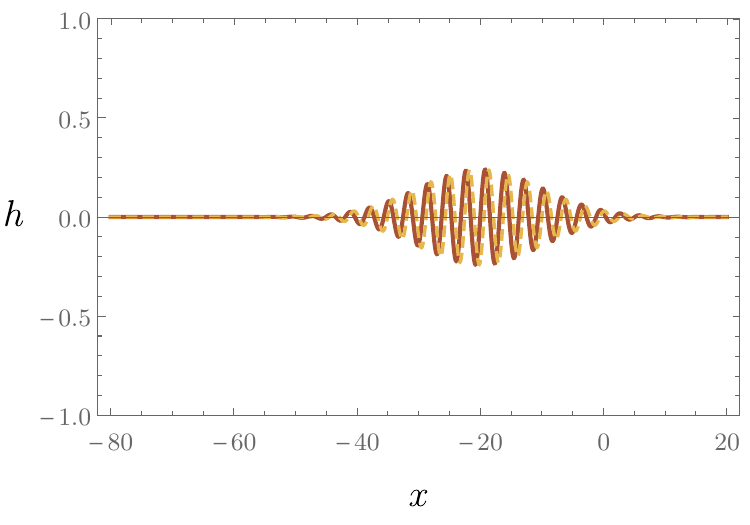}
        \caption{Evolved state at $\eta=\eta_0+3\sigma$.}
        \label{Fig:sub4d}
    \end{subfigure}
    \hfill
    \begin{subfigure}[b]{0.31\textwidth}
        \includegraphics[width=\columnwidth]{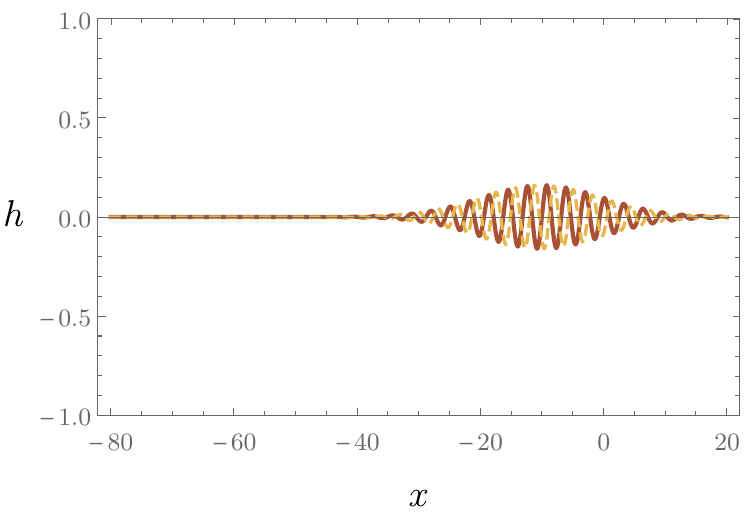}
        \caption{Evolved state at $\eta=\eta_0+4\sigma$.}
        \label{Fig:sub4e}
    \end{subfigure}
    \hfill
    \begin{subfigure}[b]{0.31\textwidth}
        \includegraphics[width=\columnwidth]{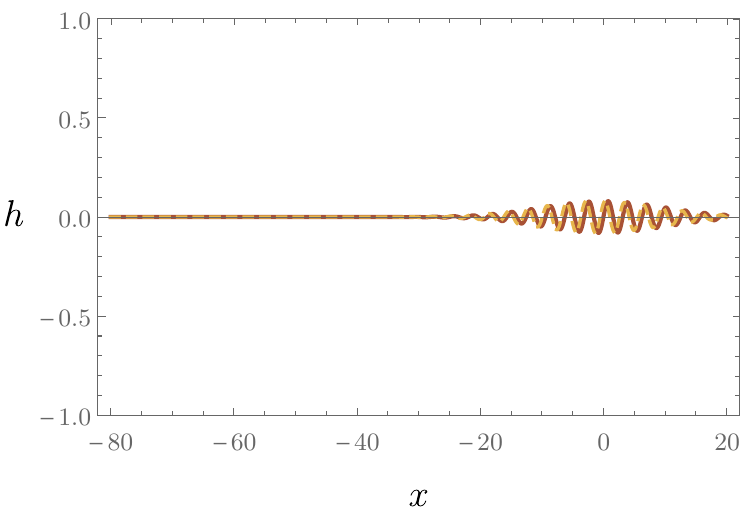}
        \caption{Evolved state at $\eta=\eta_0+5\sigma$.}
        \label{Fig:sub4f}
    \end{subfigure}
    \vspace{-0.5cm}
\caption{Evolution of the massless $h_1$ (continuous red curve) and massive $h_2$ (dashed yellow curve) components of the solution \eqref{Eq:wave-packet-sol_smallmass} in the `temporally unresolved' case, with the same initial conditions, parameters, and time spacing used in Figure~\ref{Fig:wavepackets_smallmass}. The two modes evolve independently, and both experience the usual amplitude damping due to the expansion of the Universe. Moreover, the massive mode experiences dispersion. The separation between the two components of the signal, due to the subluminal group velocity of the massive one, is particularly evident in the bottom panels.}\label{Fig:wavepackets_smallmass_eigenstates}
\end{figure}

The evolution of the wave packets is illustrated in Figures~\ref{Fig:wavepackets_smallmass} and \ref{Fig:wavepackets_smallmass_eigenstates}.
The group velocity and the phase velocity of the $h_2$ mode are given, respectively, by\footnote{These results crucially depend on the fact that we have not approximated the phase function \eqref{Eq:phase_function} at the outset. In fact, using the approximations~\eqref{Eq:dS-sol-small-mass} to build the wave packets would lead to the incorrect prediction of a superluminal group velocity.} 
\begin{subequations}
\begin{align}
{\rm v}_{g}&=-\frac{\pa W}{\pa \eta}\left(\frac{\pa W}{\pa x}\right)^{-1}\Bigg\lvert_{W=0}=\frac{1}{\sqrt{1+\frac{|\xi|^2}{k_0^2\eta^2}}}\approx 1-\frac{|\xi|^2}{2k_0^2\eta^2}+{\cal O}\left(\frac{|\xi|}{k_0|\eta|}\right)^4~, \\
{\rm v}_{p}&=-\frac{\pa \widetilde{W}}{\pa \eta}\left(\frac{\pa \widetilde{W}}{\pa x}\right)^{-1}\Bigg\lvert_{\widetilde{W}={\rm const.}}=\sqrt{1+\frac{|\xi|^2}{k_0^2\eta^2}}\approx 1+\frac{|\xi|^2}{2k_0^2\eta^2}+{\cal O}\left(\frac{|\xi|}{k_0|\eta|}\right)^4~.
\end{align}
\end{subequations}
We observe that ${\rm v}_g<1$ at all times, whereas the phase velocity ${\rm v}_p$ is superluminal. In both cases, deviations from the speed of light are small, since in this regime we are assuming $|\xi|\ll k_0|\eta|$, but they increase with time as $|\eta|$ decreases towards the future.

We can use the above results to compute the difference in arrival time $\Delta T$ between the massless and massive components, considering a propagation distance $L$ between the source and an observer at redshift $z=0$ (corresponding to $\eta=-1/H$)
\be
\Delta T=\frac{L}{{\rm v}_g}-L\approx \frac{|\xi|^2H^2}{2k_0^2}L\approx\frac{m_{\rm FP}^2}{2k_0^2}L~.
\ee
Since the wave packets have width $\sim\sigma\gg1/k_0$, the condition that ensures they can be resolved as separate signals is that $\Delta T\gg \sigma$. That is, the source must be at a distance
\be\label{Eq:separation_threshold}
L\gg L_{\rm res.}\equiv\sigma\frac{ 2k_0^2}{m_{\rm FP}^2}\sim \left(\frac{\sigma}{0.1\,{\rm s}}\right) \left(\frac{k_0}{100\, {\rm Hz}} \right)^2\left(\frac{10^{-22}\, {\rm eV}}{m_{\rm FP}} \right)^{2} {\rm Gpc} ~.
\ee
In terms of redshift, this condition reads $(1+z_0)\gg \sigma \frac{2H k_0^2}{m_{\rm FP}^2}$, for a source at $z_0$.
The quantity on the right-hand side of \eqref{Eq:separation_threshold} has been dubbed in the literature `coherence length', suggesting that after separation one enters a `decoherent regime' \cite{Max:2017kdc}. However, as we show in Section~\ref{Sec:Coherence}, the massive and massless components of the signal can retain their coherence also after they spatially separate. Such terminology is therefore misleading. For this reason, we refer to the two regimes discussed above as  `temporally resolved' and `temporally unresolved', respectively, depending on whether condition \eqref{Eq:separation_threshold} is satisfied or not.

Dispersion effects due to a nonzero mass also affect the temporal profile of the $h_2$ mode. This can be easily seen in the example above. The Gaussian envelope $e^{-\frac{W^2}{2\sigma^2}}$ of the wave packet \eqref{Eq:h2solution_smallmass} is peaked on the trajectory $W(\eta,x)=0$. Since $W$ is a nonlinear function of time, also the temporal width of the signal depends on time (while its spatial profile is unaltered). Specifically, by Taylor-expanding the argument of the envelope function to second order around the peak trajectory, one obtains that the variance of the time profile of the signal, to leading order in $\xi$, is $\sim\left(1+\frac{\xi^2}{k_0^2\eta_p^2}\right)\sigma^2$, where $\eta_p$ is the position of the peak. A third-order expansion reveals that the temporal profile is skewed in the direction of increasing time, with skewness $\sim\sigma\xi^2/(k_0^2|\eta_p|^3)$.

In summary, in the `temporally resolved' regime, the delayed copies (or `echoes') of the gravitational wave signal carried by the $h_2$ mode undergo two main effects: a distortion of their temporal profile due to the mass $m_{\rm FP}$, and amplitude suppression by a factor $\sin^2\theta$. The analysis of the propagation of wave packets given above also clarifies the different regimes of applicability of the luminosity distance formulas, \eqref{Eq:GW_luminositydistance} and \eqref{Eq:luminositydistance_resolvedregime}.  We remark that the derivation of Eq.~\eqref{Eq:GW_luminositydistance} only holds in the `temporally unresolved' regime, where the massless and massive signal components are overlapping.  However, for high redshift sources, such that condition \eqref{Eq:separation_threshold} holds and one enters the `temporally resolved' regime, the massless component is observed first at the detector and as a separate signal component, in which case the luminosity distance is given by Eq.~\eqref{Eq:luminositydistance_resolvedregime}.

\subsubsection{Regime 2: nonpropagating massive graviton ($m_{\rm FP}/H\gg |\eta|k$)}
In this regime, defined by $m_{\rm FP}/H\gg |\eta|k$\,, or equivalently $|\xi|\gg|\eta|k$\,, the solution \eqref{Eq:dS-masseigenstates-sol-v-approx} behaves as
\begin{equation}
        v_{\textbf{k}}(\eta)\approx \frac{B_{\textbf{k}}\sqrt{-\eta}}{H\sqrt{|\xi|}}\cos\left(|\xi|+|\xi|\ln\left(\frac{-k\eta}{2|\xi|}\right)+\phi_{\textbf{k}}\right)~,\quad |\xi|\gg|\eta|k~,
    \end{equation}
which describes a nonpropagating field. Note, however, that we are not necessarily constrained to super-horizon scales in this regime: for values of $m_{\rm FP}$ much greater than the Higuchi lower bound, one may well have $|\eta|k\gg1$, corresponding to sub-horizon scales whereby the massless mode freely oscillates.  In terms of the original metric perturbations, we obtain
\begin{subequations}\label{Eq:dS-sol-large-mass}
\begin{align}
\begin{split}
    h_{\textbf{k}}(z)&\approx A_{\textbf{k}}\cos\theta\,\left(\cos\left(\frac{k}{H}(1+z)+\varphi_{\textbf{k}}\right)+\frac{k}{H}(1+z)\sin\left(\frac{k}{H}(1+z)+\varphi_{\textbf{k}}\right)\right)\\
    &\quad -\doublewidetilde{B}_{\textbf{k}}\sin\theta \,(1+z)^{3/2}\cos\left(|\xi|+|\xi|\ln\left(\frac{k(1+z)}{2H|\xi|}\right)+\phi_{\textbf{k}}\right) ~,
\end{split}\label{Eq:dS-sol-large-mass-1}\\
\begin{split}
    \tilde{h}_{\textbf{k}}(z) & \approx A_{\textbf{k}}\cos\theta\,  \left(\cos\left(\frac{k}{H}(1+z)+\varphi_{\textbf{k}}\right)+\frac{k}{H}(1+z)\sin\left(\frac{k}{H}(1+z)+\varphi_{\textbf{k}}\right)\right)\\
    & \quad + \doublewidetilde{B}_{\textbf{k}}\frac{\cos^2\theta}{\sin\theta}(1+z)^{3/2}\cos\left(|\xi|+|\xi|\ln\left(\frac{k(1+z)}{2H|\xi|}\right)+\phi_{\textbf{k}}\right)~.\label{Eq:dS-sol-large-mass-2}
\end{split}
\end{align}
\end{subequations}
Note the fast superimposed log-periodic oscillations on top of the standard GR-like oscillatory solution. This is a distinctive feature of bigravity in the large-mass regime in momentum space for a de Sitter background, and is associated with the nonpropagating (i.e., ultralocal) massive mode.\footnote{In contrast, this feature is absent in Minkowski, where the Fourier components of the massive mode in the nonpropagating regime oscillate with constant frequency $m_{\rm FP}$.} However, such superimposed oscillations are not associated with a propagating degree of freedom and are therefore not directly observable by a distant observer. We also note that the amplitude of $h_{2}$ scales as $\sim a^{-3/2}$, and therefore its energy density scales as that of nonrelativistic matter $\sim a^{-3}$.

\vspace{0.3cm}

\underline{Gravitational-wave luminosity distance}

\vspace{0.2cm}

We can compute the luminosity distance in this regime following similar steps to those in the previous subsection. We impose initial conditions on $\tilde{h}$ such that this mode is initially unexcited, i.e.,~$\tilde{h}_{\textbf{k}}(z_0)=\tilde{h}_{\textbf{k}}^{\prime}(z_0)=0$ for all $\textbf{k}$, where $z_0$ is the redshift at emission. As before, we also impose initial conditions $h_{\textbf{k}}(z_0) = h^{\rm GR}_{\textbf{k}}(z_0)=f_{\textbf{k}}$, to facilitate the comparison with propagation in GR. We only consider positive frequency mode functions for the propagating component in $h_{\textbf{k}}$, but we include both independent mode functions for the nonpropagating massive component. By construction, the GR solution is the same as in Eq.~\eqref{Eq:GRsol_lumdis1}, while the bimetric solution in this case reads
\be\label{Eq:HRsol_lumdis2}
\begin{split}
h_{\textbf{k}}(z)& =f_{\textbf{k}}\Bigg\{\cos^2\theta\, e^{i\frac{k}{H}(z-z_0)}\left(\frac{H-ik(1+z)}{H-ik(1+z_0)}\right)+\sin^2\theta\left(\frac{1+z}{1+z_0}\right)^{3/2}\times\\
& \quad\left[\cos\left(|\xi|\ln\left(\frac{1+z}{1+z_0}\right) \right) -\frac{1}{2|\xi|}\left(1-\frac{2ik}{H}(1+z_0)+\frac{2H}{H-ik(1+z_0)} \right)\right]\sin\left(|\xi|\ln\left(\frac{1+z}{1+z_0}\right) \right) \Bigg\}~,
\end{split}
\ee
\be\label{Eq:HRsol_lumdis2tilde}
\begin{split}
\tilde{h}_{\textbf{k}}(z)&=f_{\textbf{k}}\cos^2\theta\Bigg\{e^{i\frac{k}{H}(z-z_0)}\left(\frac{H-ik(1+z)}{H-ik(1+z_0)}\right)-\left(\frac{1+z}{1+z_0}\right)^{3/2}\times\\
& \quad\left[\cos\left(|\xi|\ln\left(\frac{1+z}{1+z_0}\right) \right) -\frac{1}{2|\xi|}\left(1-\frac{2ik}{H}(1+z_0)+\frac{2H}{H-ik(1+z_0)} \right)\right]\sin\left(|\xi|\ln\left(\frac{1+z}{1+z_0}\right) \right) \Bigg\}\,.
\end{split}
\ee
The second term in Eq.~\eqref{Eq:HRsol_lumdis2} is nonpropagating, therefore it is unobservable for an observer at redshift $z=0$.  Hence, the only effect directly relevant for observations is the $\cos^2\theta$ suppression factor compared to GR in the propagating component of the bimetric solution in \eqref{Eq:HRsol_lumdis2}. This is similar to the `temporally resolved' sub-regime considered in Section~\ref{Sec:regime1}; however, in this case, there are no `echoes' of the original signal carried by the massive graviton. Therefore, the gravitational-wave luminosity distance in this regime is once again given by Eq.~\eqref{Eq:luminositydistance_resolvedregime}.

\vspace{0.3cm}

\underline{Propagation of wave packets}

\vspace{0.2cm}

\begin{figure}[h]
    \centering
     \begin{subfigure}[b]{0.31\textwidth}
    \includegraphics[width=\columnwidth]{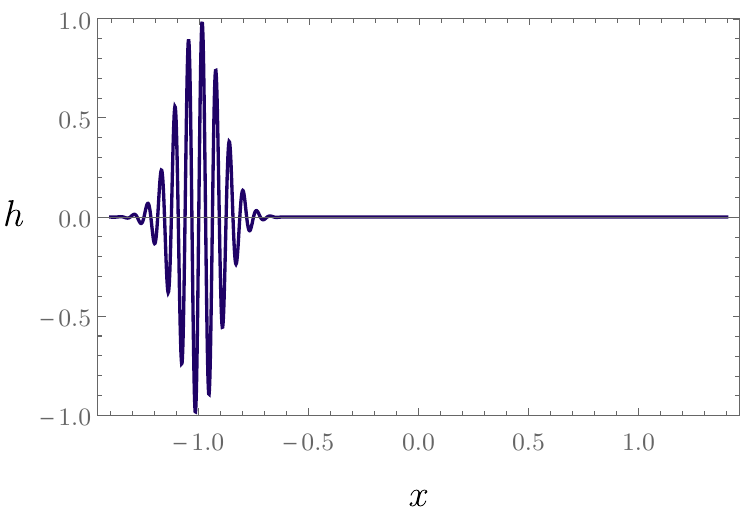}
    \caption{Initial state at $\eta=\eta_0$.}\label{Fig:sub5a}
    \end{subfigure}
      \hfill
 \begin{subfigure}[b]{0.31\textwidth}
    \includegraphics[width=\columnwidth]{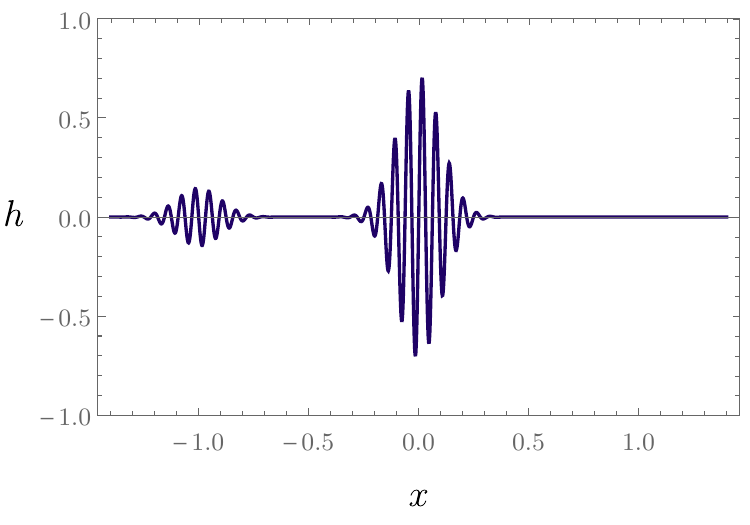}
    \caption{Evolved state at $\eta=\eta_0-x_0$.}\label{Fig:sub5b}
    \end{subfigure}
      \hfill
    \begin{subfigure}[b]{0.31\textwidth}
    \includegraphics[width=\columnwidth]{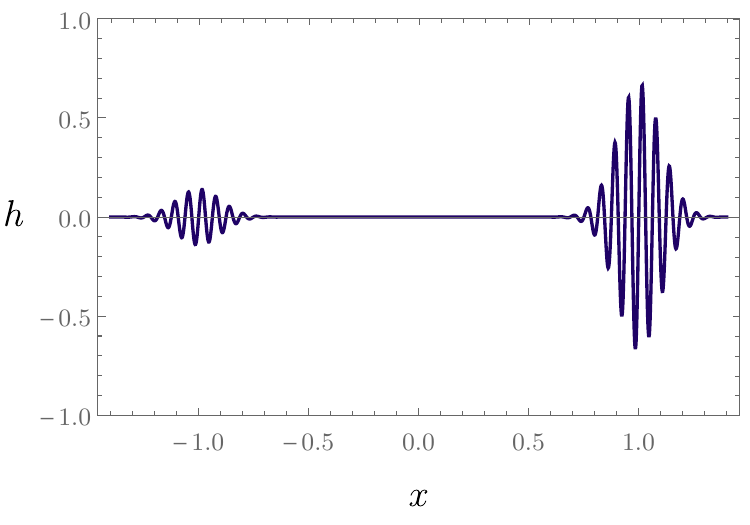}
    \caption{Evolved state at $\eta=\eta_0-2x_0$.}\label{Fig:sub5c}
    \end{subfigure}
    \vspace{0.5cm} 
    \begin{subfigure}[b]{0.31\textwidth}
    \includegraphics[width=\columnwidth]{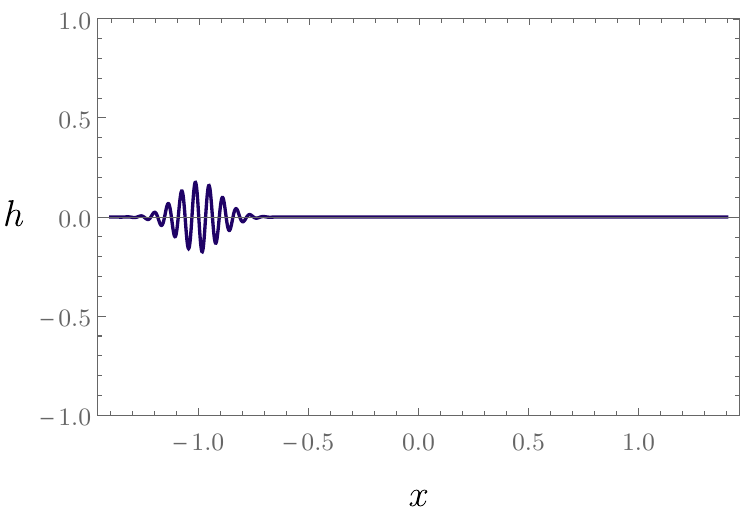}
        \caption{Evolved state at $\eta=\eta_0-3x_0$.}
        \label{Fig:sub5d}
    \end{subfigure}
    \hfill
    \begin{subfigure}[b]{0.31\textwidth}
        \includegraphics[width=\columnwidth]{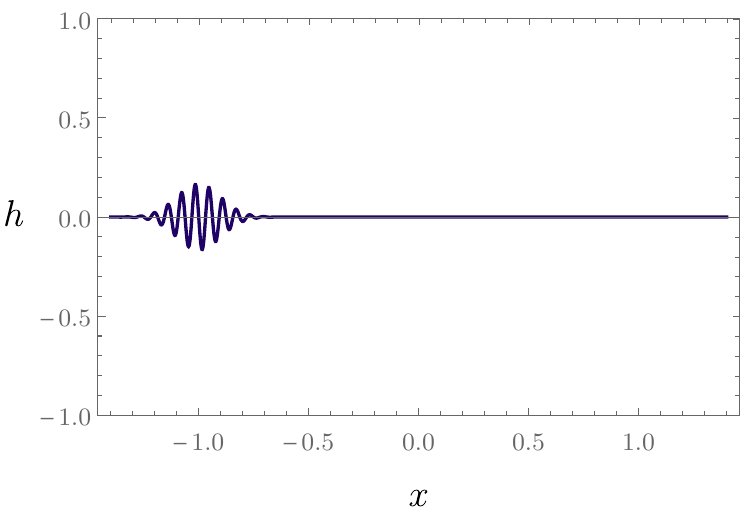}
        \caption{Evolved state at $\eta=\eta_0-4x_0$.}
        \label{Fig:sub5e}
    \end{subfigure}
    \hfill
    \begin{subfigure}[b]{0.31\textwidth}
        \includegraphics[width=\columnwidth]{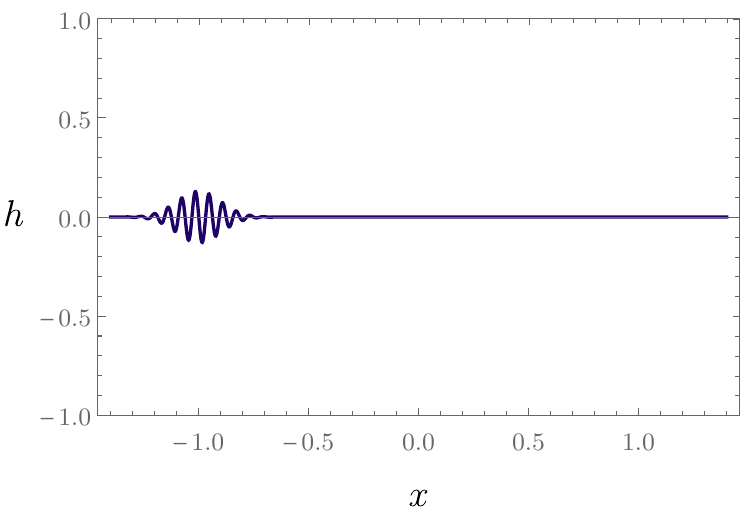}
        \caption{Evolved state at $\eta=\eta_0-5x_0$.}
        \label{Fig:sub5f}
    \end{subfigure}
    \vspace{-0.5cm}
\caption{The figure depicts the evolution of a wave packet for the $h$ mode in the `nonpropagating regime', for initial data such that $\tilde{h}(\eta_0,x)=\tilde{h}^{\prime}(\eta_0,x)=0$, corresponding to the analytical solution \eqref{Eq:wave-packet-sol_largemass-h1}-\eqref{Eq:wave-packet-sol_largemass-h2}, with parameters $|\xi|=10^5$, $\theta=\pi/6$, $H=10^{-1}$, $k_0=10^3H$, $\sigma=10^{-1}$, $x_0=-1$, $\eta_0=-20$,   $N=\sigma^2/(4|\eta_0|)$. The massless tensor mode propagates as in GR, the only difference being a constant $\cos^2\theta$ suppressing factor in the amplitude. The massive tensor mode remains localized around the emission site, its amplitude is suppressed by a $\sin^2\theta$ factor, and it features log-periodic oscillations in time as well as a time-dependence of the amplitude $\sim |\eta|^{3/2}$. The plots above are equally spaced in conformal time, starting at $\eta=\eta_0$ and increasing in time steps $\Delta\eta=-x_0$. }\label{Fig:wavepackets_largemass}
\end{figure}

To illustrate the dynamics in this regime, we now consider a one-dimensional example subject to the initial conditions specified above, assuming that the massless eigenstate $u$ propagates along the positive $x$ direction. We assume the same profile for $f_{\textbf{k}}$ as specified in Eq.~\eqref{Eq:initialconditionAk} for the example in the previous subsection. Also in this case, it is convenient to re-express the time-dependence in terms of conformal time instead of redshift. To obtain the wave packets in real space, we compute the real part of the Fourier integrals $h(\eta,x)=\Re\left[\frac{1}{2\pi}\int_{-\infty}^{+\infty}\de k\;h_{\textbf{k}}(\eta) e^{ik x}\right]$ and $\tilde{h}(\eta,x)=\Re\left[\frac{1}{2\pi}\int_{-\infty}^{+\infty}\de k\;\tilde{h}_{\textbf{k}}(\eta) e^{ik x}\right]$. Finally, the solution can be decomposed as in \eqref{Eq:wave-packet-sol_smallmass}, as a superposition of a propagating massless wave $h_1$ and a nonpropagating  massive mode $h_2$, where the two components now read as
\be\label{Eq:wave-packet-sol_largemass-h1}
h_{1}(\eta,x)=\frac{N\, e^{-\frac{U^2}{2\sigma^2}}}{\sqrt{2\pi}\sigma}\Big[ \left(1+\frac{\eta\, U}{\sigma^2}\right)\cos\left(k_0 U\right)+k_0\eta\sin\left(k_0 U\right)\Big]~,
\ee
with $U=\eta-x-\eta_0+x_0$, and
\be\label{Eq:wave-packet-sol_largemass-h2}
\begin{split}
&h_{2}(\eta,x)=\\
&\frac{N\,e^{-\frac{(x-x_0)^2}{2\sigma^2}}}{\sqrt{2\pi}\sigma}\left(\frac{\eta}{\eta_0}\right)^{3/2}\Bigg\{\cos\left(|\xi|\ln\left(\frac{\eta}{\eta_0}\right)\right)\left[\left(1-\frac{\eta_0}{\sigma^2}(x-x_0) \right)\cos\big(k_0(x-x_0)\big)-k_0\eta_0\sin\big(k_0(x-x_0)\big)\right] \\
&-\frac{1}{2|\xi|}\sin\left(|\xi|\ln\left(\frac{\eta}{\eta_0}\right)\right)
\left[\left(3-2k_0^2\eta_0^2-\frac{\eta_0}{\sigma^2}\big(3(x-x_0)+2\eta_0\big)+\frac{2\eta_0^2}{\sigma^4}(x-x_0)^2 \right)\cos\big(k_0(x-x_0)\big)  \right. \\
&-\left. \left(3k_0\eta_0-\frac{4\eta_0^2}{\sigma^2}k_0(x-x_0)\right)\sin\big(k_0(x-x_0)\big) \right]
\Bigg\}~.
\end{split}
\ee
Note that the solution for $h_1$ in Eq.~\eqref{Eq:wave-packet-sol_largemass-h1} coincides with that obtained in Eq.~\eqref{Eq:wave-packet-sol_smallmass-h1} for the propagating regime. This is a direct consequence of the fact that the solution for the massless mode $u$ is exact and therefore independent of the regime under consideration. By contrast, the massive mode $v$ has different asymptotics in each regime, and hence only $h_2$ is regime-dependent. The evolution of the $h$ mode in this example is illustrated in Figure~\ref{Fig:wavepackets_largemass}.

\section{Decoherence of bigravity oscillations?}\label{Sec:Coherence}

Regarding the detection of tensor modes in the bigravity framework, only $h$ is directly relevant, as it is the only one that couples to matter. In previous sections, we showed that $h$ propagates as a superposition of mass eigenstates: one massless, $u$, and one massive, $v$. These modes obey different dispersion relations and, as a result, the corresponding signals in real space propagate with different group velocities. In turn, this implies that, for signals with finite duration (such as wave packets) and sufficiently large propagation trajectories, there comes a point where the two signals no longer overlap and, from that point on, propagate without displaying any interference patterns. This phenomenon has been previously referred to in the literature as `decoherence of bigravity oscillations' \cite{Max:2017flc, Max:2017kdc}. However, in the absence of matter fields, since the system is purely classical, there is nothing that could cause two initially coherent wave packets to decohere.\footnote{Here we adopt the standard definition of spatiotemporal coherence for classical radiation, as used also in optics \cite{Loudon:2000}.} The purpose of this section is to show that, even in the presence of incoherent matter sources coupled to gravity, wave packets of the mass eigenstates can preserve their initial coherence even long after they separate. Moreover, unlike neutrino oscillations, correlation functions of mass eigenstates decay algebraically, and therefore, there is no well-defined notion of a coherence length associated with bigravity oscillations, contrary to previous claims.

\subsection{Wave packets propagating in an incoherent medium}

For simplicity, let us consider in this section a flat Minkowski background ($a=1$, $c=1$, $y=y_*$, and thus $b=y_*$), which is a reasonable approximation also in the late universe for distances and time scales much smaller than the Hubble radius. We use the same notation for the time coordinate $\eta$ as in previous section; however, in this case the range of $\eta$ is the full real line. This will allow us to compute correlation functions of mass eigenstates in real space analytically. For a Minkowski background, the system of equations \eqref{Eq:MSlike-system} above boils down to
\begin{subequations}
\begin{align}
&\mu_{\textbf{k}}^{\prime\prime}+ k^2 \mu_{\textbf{k}}+m^2 \lambda(y_*) \left(\mu_{\textbf{k}}-y_*^{-1}\tilde{\mu}_{\textbf{k}} \right)=\frac{2}{M_g^2}\pi_{\textbf{k}}~,\\
&\tilde{\mu}_{\textbf{k}}^{\prime\prime} + k^2\tilde{\mu}_{\textbf{k}}+\frac{m^2}{\alpha^2} y_*^{-2}\lambda(y_*) \left(\tilde{\mu}_{\textbf{k}}-y_*\mu_{\textbf{k}} \right)=0~.
\end{align}
\end{subequations}
Following similar steps as in Section~\ref{Sec:TensorModes_dS}, we obtain the propagation equations for the mass eigenstates
\begin{subequations}\label{Eq:noisy_system}
\begin{align}
&u_{\textbf{k}}^{\prime\prime}+ k^2 u_{\textbf{k}}=\frac{2\cos\theta}{M_g^2}\pi_{\textbf{k}}~,\\
&v_{\textbf{k}}^{\prime\prime} +\omega_k^2 v_{\textbf{k}}=-\frac{2\sin\theta}{M_g^2}\pi_{\textbf{k}}~,
\end{align}
\end{subequations}
 where $\omega_k\coloneqq \sqrt{k^2+m_{\rm FP}^2}$~. 
If we restrict our attention to positive-frequency solutions of the inhomogeneous system~\eqref{Eq:noisy_system}, the general solution can be written as
\begin{subequations}
\begin{align}
u_{\textbf{k}}(\eta)&=A_{\textbf{k}} e^{-ik(\eta-\eta_0)-i\varphi_{\textbf{k}}}+\frac{2\cos\theta}{M_g^2} \int_{\eta_0}^\eta\de\eta'\, \frac{\sin(k(\eta-\eta'))}{k}\pi_{\textbf{k}}(\eta')~,\label{Eq:noisysolution_u}\\
v_{\textbf{k}}(\eta)&=B_{\textbf{k}} e^{-i\omega_k(\eta-\eta_0)-i\phi_{\textbf{k}}}-\frac{2\sin\theta}{M_g^2} \int_{\eta_0}^\eta\de\eta'\, \frac{\sin(\omega_k(\eta-\eta'))}{\omega_k}\pi_{\textbf{k}}(\eta')~,\label{Eq:noisysolution_v}
\end{align}
\end{subequations}
where $\eta_0$ represents some initial time where we set initial conditions (for instance, corresponding to the emission event, as we did in the previous section). The Fourier components of the matter source are subject to the reality conditions $\pi_{\textbf{k}}=\pi_{-\textbf{k}}^{*}$. On the contrary, it is not necessary to impose reality conditions on the homogeneous solution, represented by the first terms in \eqref{Eq:noisysolution_u}--\eqref{Eq:noisysolution_v}, as we will take the real part of the Fourier integral to ensure reality of the wave-packet solution. For simplicity, in the following we assume propagation only in the $x$ direction, that is $\textbf{k}=k_x\hat{\textbf{x}}$.
Taking the inverse Fourier transform, we obtain the corresponding wave-packet solutions in position space
\begin{subequations}
\begin{align}
u(\eta,x)&=\Re\left[\int_{-\infty}^{+\infty} \frac{\de k_x}{2\pi}\, A_{\textbf{k}} e^{-i\left(|k_x|(\eta-\eta_0)-k_xx+\varphi_{\textbf{k}}\right)}\right]+\frac{2\cos\theta}{M_g^2} \int_{\eta_0}^\eta\de\eta' \int_{-\infty}^{+\infty} \frac{\de k_x}{2\pi}\, \frac{\sin(k_x(\eta-\eta'))}{k_x}e^{ik_x x}\pi_{\textbf{k}}(\eta')~,\\
v(\eta,x)&=\Re\left[\int_{-\infty}^{+\infty} \frac{\de k_x}{2\pi}\,B_{\textbf{k}} e^{-i\left(\omega_k(\eta-\eta_0)-k_xx+\phi_{\textbf{k}}\right)}\right]-\frac{2\sin\theta}{M_g^2} \int_{\eta_0}^\eta\de\eta' \int_{-\infty}^{+\infty} \frac{\de k_x}{2\pi}\,\frac{\sin(\omega_k(\eta-\eta'))}{\omega_k}e^{ik_xx}\pi_{\textbf{k}}(\eta')~.
\end{align}
\end{subequations}
As stated above, the $u$ mode propagates at the speed of light, consistent with its identification as the massless graviton. In contrast, the $v$ mode propagates subluminally, with group velocity determined by $\frac{d \omega_k}{d k}=k/\omega_k<1$.

Next, let us model the anisotropic stress $\pi_{\textbf{k}}(\eta)$ as a stochastic source, in particular as Gaussian white noise. This model consists of uncorrelated random variables at
every point $(\eta,\textbf{k})$, each sampled from a Gaussian distribution with zero mean and fixed variance $\noise^2$, which is a measure of noise strength. Despite its simplicity, this is a powerful tool for describing the behavior of complex stochastic systems \cite{oksendal2003sde}. Mathematically we have the expectation values
\be\label{Eq:white-noise}
\langle \pi_{\textbf{k}}(\eta) \rangle=0 ~, \quad \langle \pi_{\textbf{k}}(\eta)\pi_{\textbf{k}'}(\eta') \rangle= 2
\pi \noise^2 \delta(\eta-\eta')\delta(\textbf{k}+\textbf{k}')~.
\ee
 These properties will be exploited later on. We note that we are restricting our analysis to a single spatial coordinate; however, the generalization to include functional dependence on all three spatial directions is straightforward.

\subsection{Correlation functions of mass eigenstates}

In this section, we systematically compute the correlation functions of wave-packet solutions for the mass eigenstates. In order to assess whether decoherence takes place, it is crucial to investigate the decay properties of such correlation functions. Given two modes, $V(\eta',x')$ and $\widetilde{V}(\eta,x)$, evaluated at spacetime points $(\eta,x)$ and $(\eta',x')$, respectively, their correlation function is defined as
\begin{equation}\label{Eq:definition-correlation}
    R_{V\widetilde{V}}\left(\eta,x;\eta',x'\right) := \frac{\langle V(\eta,x)\widetilde{V}(\eta',x')\rangle - \langle V(\eta,x)\rangle \langle \widetilde{V}(\eta',x')\rangle}{\sqrt{{\rm Var}\big[V(\eta,x)\big]\,{\rm Var}\big[\widetilde{V}(\eta',x')\big]}}~,
\end{equation}
where $\langle V(\eta,x)\rangle$ denotes the expectation value of $V(\eta,x)$, and ${\rm Var}\big[V(\eta,x)\big]:=\langle V^2(\eta,x)\rangle-\langle V(\eta,x)\rangle^2 $
its variance. The correlation of a single mode at two different spacetime points is known as the \textit{autocorrelation function}.

In the following, we present the results for the correlation functions of the mass eigenstates, with all intermediate steps of the calculations detailed in Appendix~\ref{App:integrals}.
For convenience, we define 
\begin{equation}
    \Delta x:=x-x'~,
\end{equation} 
since, as will be made explicit below, the spatial positions $x$ and $x'$ only appear in such combination in the expressions for the different correlation functions.

As shown below, the correlation functions between mass eigenstates always decay algebraically in the limit of large spatiotemporal separations. This shows that the system does not undergo decoherence, which is associated with an exponential decay of the correlation functions.

\subsubsection{Autocorrelation of the massless mode $u$}

\begin{figure}
    \centering    \includegraphics[width=0.6\linewidth]{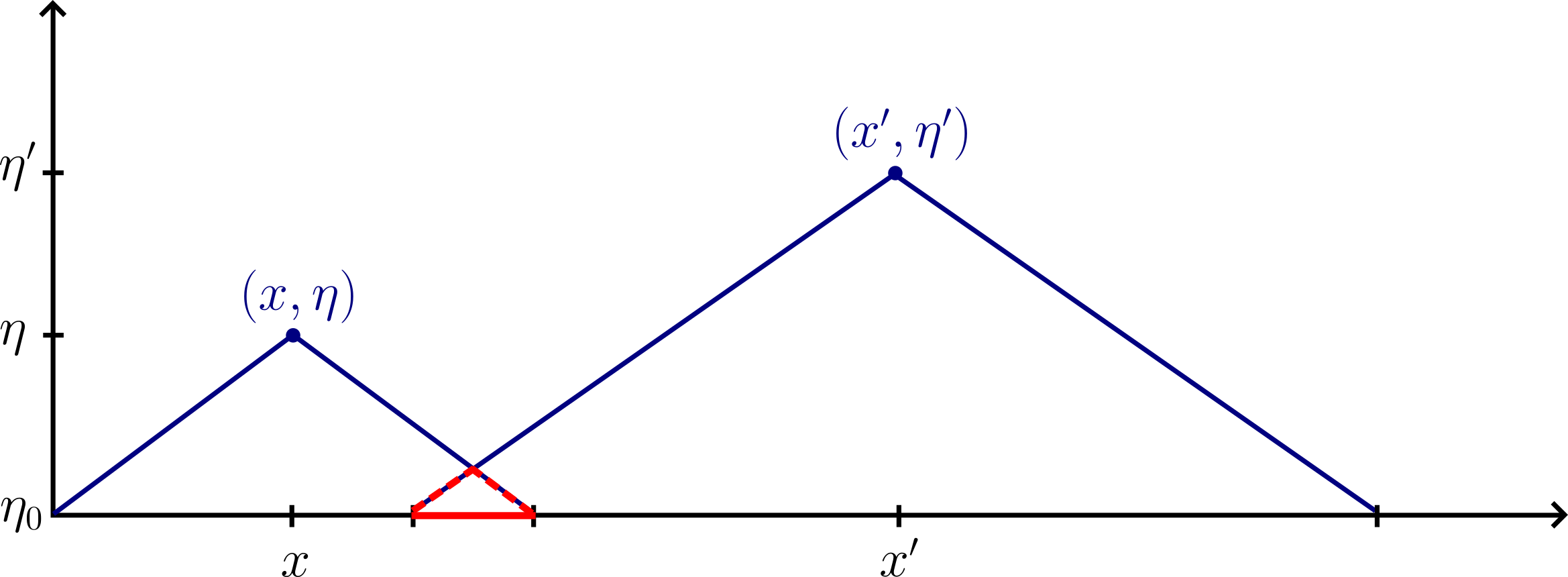}
    \caption{Illustration of the causal structure underlying the nonzero correlation between spacelike-separated events. The points $(x,\eta)$ and $(x',\eta')$ are spacelike separated but share a common causal past: both are connected to the initial time slice $\eta_0$ through their respective past lightcones (solid blue lines). The intersection of these lightcones (red dashed lines) defines a region from which signals can influence both events.}
    \label{fig:spacelike}
\end{figure}

The general expression for the autocorrelation of $u$ is given in Eq.~\eqref{Eq:autocorrelation_u} in Appendix~\ref{App:autocorrelation-u}.
This expression simplifies considerably when we consider spacetime intervals with given causality properties. For lightlike ($|\eta-\eta'|=|x-x'|$) and timelike ($|\eta-\eta'|>|x-x'|$) separated events, the autocorrelation of $u$ reduces to
\begin{equation}
    \text{R}_{uu}\big(\eta,x;\eta',x'\big)=\Theta(\eta-\eta')\frac{\eta'-\eta_0}{\eta-\eta_0}+\Theta(\eta'-\eta)\frac{\eta-\eta_0}{\eta'-\eta_0}~,
\end{equation}
where $\Theta(\eta-\eta')$ is the Heaviside function, with the convention
$\Theta(z)=0$ for $z<0$, $\Theta(z)=1$ for $z>0$, and $\Theta(0)=1/2$. Outside the lightcone $|x-x'|>|\eta-\eta'|$ we obtain
\begin{equation}
    \text{R}_{uu}\big(\eta,x;\eta',x'\big)=\left\{\begin{array}{cc}
         \frac{(\eta+\eta'-2\eta_0-|\Delta x|)^2}{4(\eta-\eta_0)(\eta'-\eta_0)}& \text{for }  |x-x'|\leq \eta+\eta'-2\eta_0~,\\
       0  & \text{for }  |x-x'|>\eta+\eta'-2\eta_0 ~.
    \end{array}\right.
\end{equation}
The reason why the correlation between spacelike-separated events can be nonzero is that, for $|x-x'|\leq \eta+\eta'-2\eta_0$, both events are causally connected to a common past event (see Figure~\ref{fig:spacelike}).

Thus, we conclude that the autocorrelation of $u$ is independent of the noise strength $\noise$, and it decays algebraically for large separations between any pair of events $(\eta,x)$ and $(\eta',x')$ (that is, for either large spatial distances $|x-x'|$ or large temporal intervals $|\eta-\eta'|$).

\subsubsection{Autocorrelation of the massive mode  $v$}\label{Sec:autocorrelation-v}

As presented in detail in Appendix \ref{App:autocorrelation-v}, the computation of the autocorrelation of $v$ is not as straightforward as in the case of $u$, since the integrals involved require more sophisticated manipulations.

For timelike separated events, $|\eta-\eta'|>|x-x'|$, the asymptotics of the  autocorrelation function of $v$ for large separations between events is given by
\begin{equation}
    \begin{split}
         \text{R}_{vv}\big(\eta,x;\eta',x'\big) \approx & \,\, \Theta(\eta-\eta')\frac{\sqrt{2(\eta'-\eta_0)}\left((\eta-\eta')^2-\Delta x^2\right)^{1/4}\cos{\left(\frac{\pi}{4}+m_{\rm FP}\sqrt{(\eta-\eta')^2-\Delta x^2}\right)}}{\sqrt{\pi\, m_{\rm FP}(\eta-\eta_0)}(\eta-\eta')}\\
             & -\Theta(\eta'-\eta)\frac{\sqrt{2(\eta-\eta_0)}\left((\eta-\eta')^2-\Delta x^2\right)^{1/4}\cos{\left(\frac{\pi}{4}+m_{\rm FP}\sqrt{(\eta-\eta')^2-\Delta x^2}\right)}}{\sqrt{\pi\, m_{\rm FP}(\eta'-\eta_0)}(\eta-\eta')}~.
    \end{split}
\end{equation}
The equivalent expression for spacelike or lightlike separated events with a common causal past, i.e., $|\eta-\eta'|\leq|x-x'|<\eta+\eta'-2\eta_0$, is
\begin{equation}
        \text{R}_{vv}\big(\eta,x;\eta',x'\big) \approx  -\frac{\left((\eta+\eta'-2\eta_0)^2-\Delta x^2\right)^{3/4}\sin{\left(\frac{\pi}{4}+m_{\rm FP}\sqrt{(\eta+\eta'-2\eta_0)^2-\Delta x^2}\right)}}{\sqrt{2\pi \,m_{\rm FP}^3(\eta-\eta_0)(\eta'-\eta_0)}(\eta+\eta'-2\eta_0)^2}~,
\end{equation}
whereas for spacelike separated events without a common causal past, that is, $|x-x'|\geq\eta+\eta'-2\eta_0>|\eta-\eta'|$, as expected, the autocorrelation is identically zero. Therefore, as can be explicitly seen in these expressions, the autocorrelation of $v$ also decays algebraically for large separations between the events $(\eta,x)$ and $(\eta',x')$.

It is worth highlighting the difference in the behavior of the correlation function for the massless mode $u$ and the massive mode $v$, depending on the causal properties of the interval between events. In the massless case, lightlike and timelike separations form a continuous class, as events can be connected by propagation at or below the speed of light. In contrast, in the massive case, lightlike events cannot be connected by a causal trajectory and,  from the perspective of the correlation function, these behave similarly to spacelike ones. This explains why lightlike and spacelike events are grouped together in the massive case, whereas in the massless case, lightlike events are grouped with timelike ones.

\subsubsection{Correlation between the massless $u$ and the massive $v$ modes}\label{Sec:correlation-u-v}

Finally, to obtain the correlation function of $u$ and $v$ one must proceed analogously to the autocorrelation of $v$, as derived in detail in Appendix~\ref{App:correlation-u-v}. 
On the one hand, for timelike separated events, $|\eta-\eta'|> |x-x'|$, the correlation function evaluates to
\begin{equation}
    \begin{split}
        \text{R}_{uv}\big(\eta,x;\eta',x'\big) \approx &\,\,  -\Theta(\eta-\eta')\frac{4\sin^2{\left(m_{\rm FP}(\eta'-\eta_0)/2\right)}}{m_{\rm FP}(\eta-\eta_0)\sqrt{m_{\rm FP}(\eta'-\eta_0)}}-\Theta(\eta'-\eta)\frac{4\sin^2{\left(m_{\rm FP}(\eta-\eta_0)\right/2)}}{m_{\rm FP}(\eta'-\eta_0)\sqrt{m_{\rm FP}(\eta-\eta_0)}}~.
    \end{split}
\end{equation}
On the other hand, for lightlike separated events, $|\eta-\eta'|=|x-x'|$, 
\begin{equation}
    \begin{split}
        \text{R}_{uv}\big(\eta,x;\eta',x'\big) \approx &\,\,  \Theta(\eta-\eta')\frac{2\cos{\left(m_{\rm FP}(\eta'-\eta_0)\right)}-1}{m_{\rm FP}(\eta-\eta_0)\sqrt{m_{\rm FP}(\eta'-\eta_0)}}+\Theta(\eta'-\eta)\frac{2\cos{\left(m_{\rm FP}(\eta-\eta_0)\right)}-1}{m_{\rm FP}(\eta'-\eta_0)\sqrt{m_{\rm FP}(\eta-\eta_0)}}~,
    \end{split}
\end{equation}
whereas for spacelike separated events with a common causal past, i.e., $|\eta-\eta'|<|x-x'|< \eta+\eta'-2\eta_0$, we obtain
\begin{equation}
    \begin{split}
        \text{R}_{uv}\big(\eta,x;\eta',x'\big) \approx &\,\,  \Theta(\eta-\eta')\frac{\sqrt{2}\,(\eta+\eta'-2\eta_0-|\Delta x|)\sin{\left(\frac{\pi}{4}+m_{\rm FP}\sqrt{(\eta'-\eta_0)^2-(\eta-\eta_0-|\Delta x|)^2}\right)}}{\sqrt{\pi}\,m_{\rm FP}^2(\eta-\eta_0-|\Delta x|)\sqrt{\eta'-\eta_0}\,[(|\Delta x|-(\eta-\eta'))(\eta+\eta'-2\eta_0-|\Delta x|)]^{1/4}}\\ 
             & +\Theta(\eta'-\eta)\frac{\sqrt{2}\,(\eta+\eta'-2\eta_0-|\Delta x|)\sin{\left(\frac{\pi}{4}+m_{\rm FP}\sqrt{(\eta-\eta_0)^2-(\eta'-\eta_0-|\Delta x|)^2}\right)}}{\sqrt{\pi}\,m_{\rm FP}^2(\eta'-\eta_0-|\Delta x|)\sqrt{\eta-\eta_0}\,[(|\Delta x|+(\eta-\eta'))(\eta+\eta'-2\eta_0-|\Delta x|)]^{1/4}}~.
    \end{split}
\end{equation}
Again, for spacelike separated events without a common causal past, that is, $|x-x'|\geq \eta+\eta'-2\eta_0>\eta-\eta'$, the correlation function is identically zero. We conclude that, like the autocorrelation functions computed in previous subsections, also the correlation function of $u$ and $v$ is independent of the noise strength $\noise$ and decays algebraically for events $(\eta,x)$ and $(\eta',x')$ with large separation.

\section{Discussion}\label{Sec:Discussion}

In this work, we have presented a comprehensive analysis of the cosmological propagation of gravitational waves in bimetric gravity during a late-time de Sitter epoch. In this theory, the dynamics of the Fourier modes of tensor perturbations is governed by a coupled system of two second-order differential equations. In the late universe, where the two background de Sitter metrics are the same up to a constant conformal rescaling, the system can be decoupled into its mass eigenstates and solved exactly. Since the tensor modes of the two metrics do not correspond to the mass eigenstates, this gives rise to a mixing phenomenon, which does not have an analogue in other gravitational theories. The mixing angle $\theta$ and the new mass scale $m_{\rm FP}$, associated with the massive graviton mode, represent two key features of gravitational wave propagation in bimetric gravity.

The solution for the massless mode is formally identical to the corresponding solution for a tensor mode in GR. The massive mode also admits an exact solution, which can be expressed in terms of Bessel functions of the first and second kind, as given in Eq.~\eqref{Eq:dS-masseigenstates-sol-v}.
Moreover, by restricting the mass to the range $m_{\rm FP}^2/H^2 > 13/4$ (which is just slightly above the Higuchi bound, $m_{\rm FP}^2/H^2 > 2$), we obtain the uniform approximation \eqref{Eq:dS-sol} for the solution in terms of elementary functions, which is
valid on all scales.

The behavior of the massive mode changes dramatically depending on the value of the mass $m_{\rm FP}$ relative to the wavenumber $k$, and we identify two limiting regimes of interest. The first regime, analyzed in Section~\ref{Sec:regime1}, is defined by the condition that the wavenumber is large compared to the mass, that is, $m_{\rm FP}/H\ll |\eta|k$, whereby the massive graviton is freely propagating. The relevant asymptotics of the mode functions are given in Eq.~\eqref{Eq:dS-sol-small-mass}. Since the group velocity of the massive mode is subluminal, and thus it propagates more slowly than the massless mode, depending on the distance traveled in position space,
one can identify two different sub-regimes, depending on whether the massive and massless modes can be temporally resolved or not. In the `temporally unresolved' sub-regime, the massive mode causes a modulation of the gravitational wave signal. In particular, we obtained the analytical formula \eqref{Eq:GW_luminositydistance} for the gravitational-wave luminosity distance versus the electromagnetic luminosity distance, $d_L^{\rm gw}/d_L^{\rm em}$, expressed as a function of the redshift of the emitting source. The behavior of this function varies significantly, depending on the value of the mass. Specifically, we identify a threshold value for the mass \eqref{Eq:threshold}, which was missed in previous work \cite{LISACosmologyWorkingGroup:2019mwx}. For sub-threshold values $d_L^{\rm gw}/d_L^{\rm em}$ is a monotonically increasing function of redshift, whereas it transitions to an oscillatory behavior for above-threshold values, as shown in Figure~\ref{Fig:dLgw}. In both cases, the upper bound is finite and determined solely by the mixing angle (except for the special case $\theta=\frac{\pi}{4}$, where said upper bound is divergent).  However, when the propagation distance is sufficiently large---as specified by the condition \eqref{Eq:separation_threshold}---the massive and massless components separate, and one enters the `temporally resolved' sub-regime. In this case, the massless and massive modes represent two distinct components of the signal. The amplitude of the massless mode is suppressed by a $\cos^2\theta$ factor compared to a GR solution subject to the same initial data, while the amplitude of the delayed `echo' carried by the massive mode is suppressed by a $\sin^2\theta$ factor. Moreover, unlike the massless mode, whose profile is undistorted, the temporal profile of the gravitational wave `echo' is broadened and distorted due to propagation. The gravitational-wave luminosity distance in this regime is determined solely by the massless component, and the value of $d_L^{\rm gw}/d_L^{\rm em}$ is a simple function of the mixing angle, as given in Eq.~\eqref{Eq:luminositydistance_resolvedregime}. We remark that the phenomenological parametrization of the luminosity distance proposed in Ref.~\cite{Finke:2021aom} does not apply to bigravity. We also showed that the independent measurements of the luminosity distance obtained with multimessenger event GW170817 can be used to place a constraint \eqref{Eq:constraint-theta-m-plane} in the parameter space of the theory $(\theta,m_{\rm FP})$, which still allows for significant deviations from GR at high redshift. The corresponding exclusion plot is shown in Figure~\ref{Fig:ExclusionPlot}.

The second regime, defined by $m_{\rm FP}/H\gg |\eta|k$, is characterized by a nonpropagating massive mode. The relevant asymptotics for the mode functions is given in Eq.~\eqref{Eq:dS-sol-large-mass}.  As in the `temporally resolved' sub-regime discussed above, also in this case the only effect that is directly relevant for gravitational wave observations is the overall suppression by a factor $\cos^2\theta$ compared to GR. Consequently, the gravitational-wave luminosity distance is still given by Eq.~\eqref{Eq:luminositydistance_resolvedregime}. However, unlike the `temporally resolved' case, no gravitational wave `echoes' are predicted in this regime since the massive mode does not propagate. This regime includes the scenario considered in Ref.~\cite{Babichev:2016bxi}, where the massive spin-2 field behaves as cold dark matter.

We remark that the `temporally resolved' and `temporally unresolved' regimes have previously been referred to in the literature as the `coherent' and `decoherent' regimes, respectively, see Refs.~\cite{Max:2017flc,Max:2017kdc}. However, we find this terminology potentially misleading and thus deliberately avoid it in our work. As explicitly shown in Section~\ref{Sec:Coherence}, the massless and massive modes remain coherent even in the regime where they can be temporally resolved.  We analytically computed the correlation and autocorrelation functions of the mass eigenstates, showing that they decay algebraically for large separations between spacetime events. Therefore, it is not possible to define a coherence length associated with bigravity oscillations. This shows the limitation of the analogy between the classical phenomenon of bigravity oscillations and neutrino oscillations (where the correlations are inherently quantum and decay exponentially as a function of the separation between wave packets of the mass eigenstates). This effect is in principle a testable prediction of the theory: if both temporally-resolved components are observed at the detector, the spatiotemporal coherence of the gravitational wave signal may be probed through the autocorrelation function of the gravitational-wave strain time series.

The analytical results obtained in our work extend previous analyses \cite{Max:2017flc,Max:2017kdc} to higher redshift values, so long as the cosmological background can be described as a de Sitter geometry. It is important to emphasize that, throughout this work, we do not require the existence of self-accelerating solutions. This possibility, though theoretically interesting, poses a strong constraint on the parameter space of the theory. Suppose we only take into account the constraints coming from the Higuchi bound, cosmological constraints, solar system constraints, and constraints based on previous analyses of gravitational wave propagation \cite{Max:2017flc,Max:2017kdc}. In that case, this leaves open three main windows for the mass parameter, as discussed in the Introduction \cite{Hogas:2022owf}. The large-mass window is compatible with a massive spin-2 field dark-matter candidate, considered in Ref.~\cite{Babichev:2016bxi}.

The possibility of strengthening these constraints using our results for gravitational wave propagation in combination with observational data should be explored in future work. While there are some regimes of the theory where deviations from GR can be appreciated already with small-redshift sources (see the right panels of Figure~\ref{Fig:dLgw}), the other regimes considered in this paper are more subtle to test and require higher-redshift sources, which will be observed by next-generation gravitational wave detectors. In particular, using our analytical results, and in light of the systematic classification of different propagation regimes provided in our work, it would be interesting to perform a Bayesian forecast analysis for LISA as well as third-generation ground-based detectors. This would generalize previous studies \cite{Saltas:2023qec,Liu:2023onj} of modified gravitational wave propagation in theories with a single graviton.

This work can be extended in several directions. In order to extend our analysis to the propagation of gravitational waves emitted by higher-redshift sources, which were active before the late-time de Sitter era, we need to consider more general cosmological solutions on the viable branch of bimetric gravity. In this more general scenario, neither the ratio of the scale factors of the two metrics nor their relative lapse is constant. As a result, the system of equations~\eqref{Eq:MSlike-system} cannot be decoupled in terms of mass eigenstates at higher redshift. To address this, appropriate techniques based on multiple-scale analysis \cite{Brizuela:2023uwt,deCesare:2025ovv} must be employed to account for the influence of the evolving cosmological background on the dynamics of tensor modes, which smoothly reduces to the de Sitter case analyzed in this paper. Another natural extension of this work would be to consider multigravity theories involving $N\geq 3$ conformally related background metrics \cite{Baldacchino:2016jsz, Wood:2024acv}, with tensor perturbations corresponding to linear combinations of one massless and $N-1$ massive gravitons. In this context, it would be interesting to identify possible mass hierarchies, classify different dynamical regimes that may arise, and study their imprints on gravitational wave signals and the luminosity distance.

\section*{Acknowledgments}
We are grateful to S.~Castrignano, R.~Oliveri, I.D.~Saltas, and A.~Tagliacozzo for helpful discussions, and to an anonymous referee for their constructive feedback.
ASO acknowledges financial support from the fellowship PIF21/237
of the UPV/EHU. MdC acknowledges support from INFN iniziativa specifica GeoSymQFT. 
This work has been supported by the Basque Government Grant
\mbox{IT1628-22} and by the Grant PID2021-123226NB-I00 (funded by
MCIN/AEI/10.13039/501100011033 and by ``ERDF A way of making Europe'').
\appendix

\section{Asymptotic expansion of the $v$ mode}\label{App:bessel}

Here we provide a detailed derivation of the asymptotic behavior given in Eq.~\eqref{Eq:dS-masseigenstates-sol-v-approx} from the exact solution in Eq.~\eqref{Eq:dS-masseigenstates-sol-v} obtained earlier in the analysis. The calculation is based on the results of Ref.~\cite{dunster2025}, where uniform asymptotic expansions for Bessel functions of imaginary order are derived. 

Let us start with the asymptotic behavior of the Bessel function of the first kind $J_{\xi}(-k\eta)$ in the case where
the order $\xi$ is such that $\xi=i|\xi|$ with $|\xi|>1$. From \textsc{Theorem} 3.1 in Ref.~\cite{dunster2025}, we obtain
\begin{equation}\label{Eq:besselJ-approx}
    J_{i|\xi|}(-k\eta)\approx \frac{e^{\pi|\xi|/2}}{\sqrt{2\pi|\xi|}(r^2+1)^{1/4}}\exp\left[i\left(|\xi|(r^2+1)^{1/2}-|\xi|\ln\left(\frac{1+(r^2+1)^{1/2}}{r}\right)-\frac{\pi}{4}\right)\right] ~,
\end{equation}
with $r\coloneqq-k\eta/|\xi|$. This approximation holds as long as the condition
\begin{equation}\label{Eq:condition-approximation}
   \Delta\coloneqq \frac{\left(5-3(r^2+1)^{1/4}\right)}{24|\xi|(r^2+1)^{3/4}}\ll1~,
\end{equation}
is fulfilled.
Note that the approximation improves as $|\xi|$ increases, and that \eqref{Eq:condition-approximation} holds already for $|\xi|\sim 1$.
The reason why Eq.~\eqref{Eq:besselJ-approx} is not valid for $|\xi|\leq 1$, even in the limit where $r\gg 1$, stems from the fact that, in the derivation of this formula, terms of the order $\mathcal{O}(|\xi|^{-n})$ are disregarded in the $n\to\infty$ limit.

On the other hand, in order to obtain the asymptotic of $Y_{\xi}(-k\eta)$, we make use of Eqs.~(3.17)--(3.18) of Ref.~\cite{dunster2025} and write
\begin{equation}
    Y_{i|\xi|}(-k\eta)=\Re\left[Y_{i|\xi|}(-k\eta)\right]+i\,\Im\left[Y_{i|\xi|}(-k\eta)\right]=\frac{\sinh(|\xi|\pi)}{\cosh{(|\xi|\pi)}-1}\Im\left[J_{i|\xi|}(-k\eta)\right]-\frac{i\sinh(|\xi|\pi)}{\cosh{(|\xi|\pi)}+1}\Re\left[J_{i|\xi|}(-k\eta)\right]\,.
\end{equation}
Finally, the approximate expression for the exact solution of $v$ given in \eqref{Eq:dS-masseigenstates-sol-v} is, 
\begin{equation}
\begin{split}
    v_{\textbf{k}}(\eta)\approx \frac{\sqrt{-\eta}}{H(k^2\eta^2+|\xi|^2)^{1/4}}\Bigg[ & \widetilde{C}_{\textbf{k}}\cos\left(\sqrt{k^2\eta^2+|\xi|^2}+|\xi|\ln\left(\frac{-k\eta}{|\xi|+\sqrt{k^2\eta^2+|\xi|^2}}\right)+\frac{\pi}{4}\right)\\
   &  +\widetilde{D}_{\textbf{k}}\sin\left(\sqrt{k^2\eta^2+|\xi|^2}+|\xi|\ln\left(\frac{-k\eta}{|\xi|+\sqrt{k^2\eta^2+|\xi|^2}}\right)+\frac{\pi}{4}\right)\Bigg]~,
\end{split}
\end{equation}
where
\begin{subequations}
\begin{align}
    \widetilde{C}_{\textbf{k}}&=-\frac{e^{\pi|\xi|/2}}{\sqrt{2\pi}}\coth\left(\frac{\pi|\xi|}{2}\right)\left(iC_{\textbf{k}}+\coth\left(\frac{\pi|\xi|}{2}\right)D_{\textbf{k}}\right)~,\\
    \widetilde{D}_{\textbf{k}}&=-\frac{e^{\pi|\xi|/2}}{\sqrt{2\pi}}\tanh\left(\frac{\pi|\xi|}{2}\right)\left(iD_{\textbf{k}}-\coth\left(\frac{\pi|\xi|}{2}\right)C_{\textbf{k}}\right)~.
\end{align}
\end{subequations}
Hence, after a redefinition of the constants, we can write
\begin{equation}
    v_{\textbf{k}}(\eta)\approx \frac{B_{\textbf{k}}\sqrt{-\eta}}{H(k^2\eta^2+|\xi|^2)^{1/4}}\cos\left(\sqrt{k^2\eta^2+|\xi|^2}+|\xi|\ln\left(\frac{-k\eta}{|\xi|+\sqrt{k^2\eta^2+|\xi|^2}}\right)+\phi_{\textbf{k}}\right)~.
\end{equation}

For completeness, it is also worth mentioning that, if desired, one can obtain an approximation of the exact solution \eqref{Eq:dS-masseigenstates-sol-v} for $\xi\geq0$ (real and non-negative) in the sub-horizon regime, characterized by $k|\eta| \gg 1$, which corresponds to the observable modes in the present universe. This case is going to be restricted to a very narrow set of theoretically allowed mass values due to the Higuchi bound $m_{\rm FP}^2>2H^2$, corresponding to admissible values
in the interval $0\leq \xi<1/2$. In this regime, one can use the asymptotic form of Bessel functions for fixed real and non-negative order and large arguments \cite{lebedev1965},
\begin{equation}
    J_{\xi}(-k\eta)\approx \sqrt{\frac{2}{-\pi \,k \eta }}\cos\left(-k\eta-\frac{\pi}{2}\xi-\frac{\pi}{4}\right) ~,
\end{equation}
\begin{equation}
    Y_{\xi}(-k\eta)\approx \sqrt{\frac{2}{-\pi \,k \eta }}\sin\left(-k\eta-\frac{\pi}{2}\xi-\frac{\pi}{4}\right) ~.
\end{equation}
Hence, after a suitable redefinition of the integration constants, we have
\be\label{Eq:solv_verylowmass}
v_{\textbf{k}}(\eta)=\tilde{B}_{\textbf{k}}\cos\left(-k\eta-\frac{\pi}{2}\xi-\phi_{\textbf{k}}\right)~.
\ee

\section{Details on the computation of the correlation functions}\label{App:integrals}

\subsection{Autocorrelation of $u$}\label{App:autocorrelation-u}
Let us begin by computing the autocovariance of $u$. From the properties \eqref{Eq:white-noise} of the Gaussian white noise, it reduces to
\be
\begin{split}
&\langle u(\eta,x)u(\eta',x')\rangle-\langle u(\eta,x)\rangle\langle u(\eta',x')\rangle=\\
&\left(\frac{2\cos\theta}{M_g^2}\right)^2 \int\limits_{\eta_0}^\eta\de\bar{\eta} \int\limits_{\eta_0}^{\eta'}\de\bar{\bar{\eta}} \int\limits_{-\infty}^{+\infty} \frac{\de k_x}{2\pi}\int\limits_{-\infty}^{+\infty} \frac{\de k'_x}{2\pi}\, \frac{\sin(k_x(\eta-\bar{\eta}))}{k_x}\frac{\sin(k'_x(\eta'-\bar{\bar{\eta}}))}{k'_x}e^{i(k_x x+k'_x x')}\langle\pi(k,\bar{\eta})\pi(k',\bar{\bar{\eta}})\rangle=\\
&\left(\frac{2\noise\cos\theta}{M_g^2}\right)^2 \int\limits_{\eta_0}^\eta\de\bar{\eta} \int\limits_{\eta_0}^{\eta'}\de\bar{\bar{\eta}} \int\limits_{-\infty}^{+\infty} \frac{\de k_x}{2\pi}\, \frac{\sin(k_x(\eta-\bar{\eta}))}{k_x}\frac{\sin(k_x(\eta'-\bar{\bar{\eta}}))}{k_x}e^{ik_x(x-x')}\delta(\bar{\eta}-\bar{\bar{\eta}})=\\
&\left(\frac{2\noise\cos\theta}{M_g^2}\right)^2 \Theta(\eta-\eta') \int\limits_{\eta_0}^{\eta'}\de\bar{\eta}\int\limits_{-\infty}^{+\infty} \frac{\de k_x}{2\pi}\, \frac{\sin(k_x(\eta-\bar{\eta}))\sin(k_x(\eta'-\bar{\eta}))}{k_x^2}e^{ik_x(x-x')}+(\eta\stackrel{\text{sym}}{\longleftrightarrow}\eta')=\\
& -\frac{1}{8} \left(\frac{\noise\cos\theta}{M_g^2}\right)^2  \Theta(\eta-\eta')\Big[4 (\eta'-\eta_0) \big(\left| x-x'+\eta -\eta'\right|+\left| x-x'-\eta +\eta'\right| \big)+\\
   &(\eta -\eta'+x-x')|\eta -\eta'+x-x'|-(\eta' -\eta+x-x')|\eta' -\eta+x-x'|-\\
   &(\eta -2\eta_0+\eta'+x-x')|\eta -2\eta_0+\eta'+x-x'|-(\eta -2\eta_0+\eta'+x'-x)|\eta -2 \eta_0+\eta'+x'-x|\Big]+
   (\eta\stackrel{\text{sym}}{\longleftrightarrow}\eta') ~.
\end{split}
\ee
where $\Theta(\eta-\eta')$ is the Heaviside function, with the convention
$\Theta(z)=0$ for $z<0$, $\Theta(z)=1$ for $z>0$, and $\Theta(0)=1/2$.

On the other hand, computing the variance of $u$ at a given event $(\eta,x)$ (which is needed for the normalization of the autocorrelation function) is straightforward
\be\label{Eq: variance-u}
\begin{split}
{\rm Var}\big[u(\eta,x)\big]=
\left(\frac{2\noise\cos\theta}{M_g^2}\right)^2 \int\limits_{\eta_0}^{\eta}\de\bar{\eta}\int\limits_{-\infty}^{+\infty} \frac{\de k_x}{2\pi}\, \frac{\sin^2(k_x(\eta-\bar{\eta}))}{k_x^2}=\left(\frac{\noise\cos\theta}{M_g^2}\right)^2(\eta-\eta_0)^2~.
\end{split}
\ee

After computing the autocovariance and variance of $u$, the autocorrelation follows directly from the definition of correlation function \eqref{Eq:definition-correlation},
\be\label{Eq:autocorrelation_u}
\begin{split}
&\text{R}_{uu}\big(\eta,x;\eta',x'\big)=\frac{\langle u(\eta,x)u(\eta',x')\rangle-\langle u(\eta,x)\rangle\langle u(\eta',x')\rangle}{\sqrt{{\rm Var}\big[u(\eta,x)\big]{\rm Var}\big[u(\eta',x')\big]}}=\\
&-\frac{1}{8(\eta-\eta_0)(\eta'-\eta_0)}  \Big[4 (\eta'-\eta_0) \big(\left| x-x'+\eta -\eta'\right|+\left| x-x'-\eta +\eta'\right| \big)+(\eta -\eta'+x-x')|\eta -\eta'+x-x'|\\
   &-(\eta' -\eta+x-x')|\eta' -\eta+x-x'|-(\eta -2\eta_0+\eta'+x-x')|\eta -2\eta_0+\eta'+x-x'|\\
   &-(\eta -2\eta_0+\eta'+x'-x)|\eta -2 \eta_0+\eta'+x'-x|\Big]\Theta(\eta-\eta')+
   (\eta\stackrel{\text{sym}}{\longleftrightarrow}\eta')~.
\end{split}
\ee

\subsection{Autocorrelation of $v$}\label{App:autocorrelation-v}

A similar computation can be carried through for the autocorrelation of the massive field $v$. We start by computing the autocovariance function
\be
\begin{split}
&\langle v(\eta,x)v(\eta',x')\rangle-\langle v(\eta,x)\rangle\langle v(\eta',x')\rangle=\\
&\left(\frac{2\noise\sin\theta}{M_g^2}\right)^2 \int\limits_{\eta_0}^\eta\de\bar{\eta} \int\limits_{\eta_0}^{\eta'}\de\bar{\bar{\eta}} \int\limits_{-\infty}^{+\infty} \frac{\de k_x}{2\pi}\, \frac{\sin(\omega_k(\eta-\bar{\eta}))}{\omega_k}\frac{\sin(\omega_k(\eta'-\bar{\bar{\eta}}))}{\omega_k}e^{ik_x(x-x')}\delta(\bar{\eta}-\bar{\bar{\eta}})=\\
&\left(\frac{2\noise\sin\theta}{M_g^2}\right)^2\Theta(\eta-\eta') \int\limits_{\eta_0}^{\eta'}\de\bar{\eta}\int\limits_{-\infty}^{+\infty} \frac{\de k_x}{2\pi}\, \frac{\sin(\omega_k(\eta-\bar{\eta}))\sin(\omega_k(\eta'-\bar{\eta}))}{\omega_k^2}e^{ik_x(x-x')}+(\eta\stackrel{\text{sym}}{\longleftrightarrow}\eta')=\\
&\left(\frac{2\noise\sin\theta}{M_g^2}\right)^2 \Theta(\eta-\eta') \int\limits_{-\infty}^{+\infty}\text{d}k_x\; e^{i k_x (x-x')}\Bigg(\frac{(\eta'-\eta_0)\cos{\left(\sqrt{k_x^2+m_{\rm FP}^2}(\eta-\eta')\right)}}{4\pi(k_x^2+m_{\rm FP}^2)}\\
&+\frac{\sin{\left(\sqrt{k_x^2+m_{\rm FP}^2}(\eta-\eta')\right)}-\sin{\left(\sqrt{k_x^2+m_{\rm FP}^2}(\eta+\eta'-2\eta_0)\right)}}{8\pi(k_x^2+m_{\rm FP}^2)^{3/2}}\Bigg)+(\eta\stackrel{\text{sym}}{\longleftrightarrow}\eta')~.
\end{split}
\ee
Hence, we have reduced the problem to the computation of the following integral
\begin{equation}\label{Eq:IntegralModev}
\begin{split}
     & \int\limits_{-\infty}^{+\infty}\text{d}k_x\,e^{i k_x (x-x')}\Bigg(\frac{(\eta'-\eta_0)\cos{\left(\sqrt{k_x^2+m_{\rm FP}^2}(\eta-\eta')\right)}}{4\pi(k_x^2+m_{\rm FP}^2)}\\
    & +\frac{\sin{\left(\sqrt{k_x^2+m_{\rm FP}^2}(\eta-\eta')\right)}-\sin{\left(\sqrt{k_x^2+m_{\rm FP}^2}(\eta+\eta'-2\eta_0)\right)}}{8\pi(k_x^2+m_{\rm FP}^2)^{3/2}}\Bigg)\\
    & =\frac{\eta'-\eta_0}{8\pi}\left[I_2(\Delta \eta,\Delta x)+I_2(-\Delta \eta,\Delta x)\right]-\frac{i}{16\pi}\left[I_3(\Delta \eta,\Delta x)-I_3(-\Delta \eta,\Delta x)-I_3(\widetilde{\Delta \eta},\Delta x)+I_3(-\widetilde{\Delta \eta},\Delta x)\right]~,
\end{split}
\end{equation}
where $\Delta \eta=\eta-\eta'>0$~, $\widetilde{\Delta \eta}=\eta+\eta'-2\eta_0>0$, $\Delta x=x-x'$, and we have introduced the auxiliary functions
\begin{equation}\label{Eq: definitionIn}
    I_n(p,q)=\int\limits_{-\infty}^{+\infty}\text{d}k_x\,\frac{e^{i (p\,\omega_k+q\,k_x)}}{\omega_k^n}~.
\end{equation}
In order to determine the autocorrelation of $v$, we must obtain an analytic expression for the integral \eqref{Eq:IntegralModev}. This involves computing the integrals $I_2(p,q)$ and $I_3(p,q)$ defined in Eq.\eqref{Eq: definitionIn}. Using standard complex integration techniques, it can be shown that
\begin{equation}\label{Eq:J2combination}
    I_2(p,q)+I_2(-p,q)=\left\{\begin{array}{ll}
         \dfrac{2\pi}{m_{\rm FP}}e^{-m_{\rm FP} \,|q|}\,,& \text{for }  |p|\leq |q|~,\\[0.5cm]
        \dfrac{2\pi}{m_{\rm FP}}\cosh{\left(m_{\rm FP}\,q\right)}-2\int\limits_{-m_{\rm FP}}^{m_{\rm FP}}\text{d}u\, \dfrac{e^{-u\,q}}{m_{\rm FP}^2-u^2}\sin{\left(\sqrt{m_{\rm FP}^2-u^2}|p|\right)}\,, & \text{for }  |p|>|q|~.
    \end{array}\right.
\end{equation}
In order to evaluate $I_3(p,q)$, we use the following identity
\be
\frac{1}{m_{\rm FP}}\frac{\partial I_1}{\partial m_{\rm FP}}=i p  I_2- I_3~,
\ee
where, analogously to \eqref{Eq:J2combination},
\begin{equation}\label{Eq:I1-exact}
    I_1(p,q)-I_1(-p,q)=\left\{\begin{array}{ll}
         0\,,& \text{for }  |p|\leq |q|~,\\
        2 i \int\limits_{-m_{\rm FP}}^{m_{\rm FP}}\text{d}u\, \dfrac{e^{-u\,q}}{\sqrt{m_{\rm FP}^2-u^2}}\cos{\left(\sqrt{m_{\rm FP}^2-u^2}\,p\right)}\,, & \text{for }  |p|>|q|~.
    \end{array}\right.
\end{equation}
This gives
\begin{equation}
    I_3(p,q)-I_3(-p,q)=\left\{\begin{array}{ll}
         i p \left(I_2(p,q)+I_2(-p,q)\right)=\dfrac{2\pi i\, p}{m_{\rm FP}}e^{-m_{\rm FP} \,|q|}\,, & \text{for }  |p|\leq |q|~,\\[.5cm]
        \begin{split}
          & i p \left(I_2(p,q)+I_2(-p,q)\right)+\\
          &2i\int\limits_{-m_{\rm FP}}^{m_{\rm FP}}\text{d}u\, \dfrac{e^{-u\,q}}{m_{\rm FP}^2}\left(\dfrac{q\, u\cos{\left(\sqrt{m_{\rm FP}^2-u^2}\,p\right)}}{\sqrt{m_{\rm FP}^2-u^2}}+p\sin{\left(\sqrt{m_{\rm FP}^2-u^2}\,p\right)}\right)
        \end{split}\,, & \text{for }  |p|>|q|~.
    \end{array}\right.
\end{equation}

For $|p|\gg1/m_{\rm FP}$~, the asymptotics of the integral in \eqref{Eq:J2combination} can be obtained using the stationary-phase method \cite{bender1999}, which gives
\be
\begin{split}
&\int\limits_{-m_{\rm FP}}^{m_{\rm FP}}\text{d}u\, \dfrac{e^{-u\,q}}{m_{\rm FP}^2-u^2}\sin{\left(\sqrt{m_{\rm FP}^2-u^2}|p|\right)} \approx\\
&\quad\frac{1}{m_{\rm FP}} \Bigg[\pi\red{\sinh}(m_{\rm FP}|q|) -\sqrt{\frac{2\pi}{m_{\rm FP}|p|}}\left(1-\frac{q^2}{p^2}\right)^{1/4}\cos\Big(m_{\rm FP}\sqrt{p^2-q^2}+\frac{\pi}{4}\Big)\Bigg]~,\;\; \mbox{for $|p|>|q|$\,,\; $m_{\rm FP}|p|\gg1$}~.
\end{split}
\ee
Thus, in this regime we find 
\be
I_2(p,q)+I_2(-p,q)\approx \frac{2}{m_{\rm FP}}\sqrt{\frac{2\pi}{m_{\rm FP}|p|}}\left(1-\frac{q^2}{p^2}\right)^{1/4}\cos\Big(m_{\rm FP}\sqrt{p^2-q^2}+\frac{\pi}{4}\Big)~,
\ee
Similarly, in the same large-$|p|$ limit we have
\be\label{Eq:I1-approx}
I_1(p,q)-I_1(-p,q)\approx2i\sqrt{\frac{2\pi}{m_{\rm FP}}}\left(p^2-q^2\right)^{-1/4}\sin\left(m_{\rm FP}\sqrt{p^2-q^2}+\frac{\pi}{4} \right)~,
\ee
and
\be
I_3(p,q)-I_3(-p,q)\approx 2i\sqrt{2\pi}\,m_{\rm FP}^{-5/2}p^{-2}\left(p^2-q^2\right)^{3/4}\sin\left(m_{\rm FP}\sqrt{p^2-q^2}+\frac{\pi}{4} \right)~.
\ee
Using these results, we finally obtain the following asymptotic formulas for large intervals between events:
\begin{itemize}
    \item timelike separated events, $|\eta-\eta'|>|x-x'|$~:
    \begin{equation}\label{Eq: correlation-vv-case1}
        \begin{split}
&\langle v(\eta,x)v(\eta',x')\rangle-\langle v(\eta,x)\rangle\langle v(\eta',x')\rangle\\
& \approx \left(\frac{2\noise\sin\theta}{M_g^2}\right)^2 \frac{(\eta'-\eta_0)\left((\eta-\eta')^2-\Delta x^2\right)^{1/4}\cos{\left(\frac{\pi}{4}+m_{\rm FP}\sqrt{(\eta-\eta')^2-\Delta x^2}\right)}}{2\sqrt{2\pi\, m_{\rm FP}^3}(\eta-\eta')}\Theta(\eta-\eta') \\
& \quad +\,(\eta\stackrel{\text{sym}}{\longleftrightarrow}\eta')
\end{split}
    \end{equation}
     \item spacelike or lightlike separated events with a common causal past, $|\eta-\eta'|\leq|x-x'|<\eta+\eta'-2\eta_0$~:
    \begin{equation}\label{Eq: correlation-vv-case2}
        \begin{split}
&\langle v(\eta,x)v(\eta',x')\rangle-\langle v(\eta,x)\rangle\langle v(\eta',x')\rangle \\
 &\approx -\left(\frac{2\noise\sin\theta}{M_g^2}\right)^2 \frac{\left((\eta+\eta'-2\eta_0)^2-\Delta x^2\right)^{3/4}\sin{\left(\frac{\pi}{4}+m_{\rm FP}\sqrt{(\eta+\eta'-2\eta_0)^2-\Delta x^2}\right)}}{4\sqrt{2\pi m_{\rm FP}^5}(\eta+\eta'-2\eta_0)^2}\Theta(\eta-\eta') \\
 & \quad +\,(\eta\stackrel{\text{sym}}{\longleftrightarrow}\eta')~.
\end{split}
    \end{equation}
     \item spacelike separated events without a common causal past, $|x-x'|\geq\eta+\eta'-2\eta_0>|\eta-\eta'|$~:
    \begin{equation}\label{Eq: correlation-vv-case3}
        \begin{split}
&\langle v(\eta,x)v(\eta',x')\rangle-\langle v(\eta,x)\rangle\langle v(\eta',x')\rangle=0 ~.
\end{split}
    \end{equation}
    Note that, unlike the previous two cases, this result is exact.
\end{itemize}

The variance of the random variable $v$ at a fixed event $(\eta,x)$ evaluates to
\begin{equation}\label{Eq: variance-v}
\begin{split}
   \text{Var}\big[v(\eta,x)\big] &  =  \left(\frac{2\noise\sin\theta}{M_g^2}\right)^2\int\limits_{-\infty}^{+\infty}\text{d}k_x\,\frac{2 (\eta-\eta_0)+\frac{\sin \left(2 (\eta -\eta_0) \sqrt{k_x^2+m_{\rm FP}^2}\right)}{\sqrt{k_x^2+m_{\rm FP}^2}}}{8 \pi  \left(k_x^2+m_{\rm FP}^2\right)}\approx \left(\frac{\noise\sin\theta}{M_g^2}\right)^2\frac{\eta-\eta_0}{m_{\rm FP}}~,
\end{split}
\end{equation}
for $m_{\rm FP}(\eta-\eta_0)\gg 1$. Finally, using Eqs.~\eqref{Eq: correlation-vv-case1}--\eqref{Eq: variance-v}, the autocorrelation of $v$ in each case is given by \eqref{Eq:definition-correlation}.

\subsection{Correlation between $u$ and $v$}\label{App:correlation-u-v}

The computation of the correlation function of $u$ and $v$ is also analogous. We first compute the covariance
\be
\begin{split}
&\langle u(\eta,x)v(\eta',x')\rangle-\langle u(\eta,x)\rangle\langle v(\eta',x')\rangle=\\
&-\frac{2\noise^2\sin(2\theta)}{M_g^4} \Theta(\eta-\eta') \int\limits_{\eta_0}^{\eta'}\de\bar{\eta}\int\limits_{-\infty}^{+\infty} \frac{\de k_x}{2\pi}\, \frac{\sin(k_x(\eta-\bar{\eta}))\sin(\omega_k(\eta'-\bar{\eta}))}{k_x\,\omega_k}e^{ik_x(x-x')} +(\eta\stackrel{\text{sym}}{\longleftrightarrow}\eta')\\
& =-\frac{2\noise^2\sin(2\theta)}{M_g^4} \Theta(\eta-\eta') \int\limits_{-\infty}^{+\infty} \frac{\de k_x}{2\pi m_{\rm FP}^2}\Bigg(\frac{\sin{\left(k_x(\eta-\eta')\right)-\cos{\left(\sqrt{k_x^2+m_{\rm FP}^2}(\eta'-\eta_0)\right)}\sin{\left(k_x(\eta-\eta_0)\right)}}}{k}\\
& +\frac{\cos{\left(k_x(\eta-\eta_0)\right)}\sin{\left(\sqrt{k_x^2+m_{\rm FP}^2}(\eta'-\eta_0)\right)}}{\sqrt{k_x^2+m_{\rm FP}^2}}\Bigg)e^{i k_x (x-x')} +(\eta\stackrel{\text{sym}}{\longleftrightarrow}\eta')~,
\end{split}
\ee
which leads to the following integral
\begin{equation}\label{Eq:IntegralModesuv}
\begin{split}
   &  \int\limits_{-\infty}^{+\infty} \frac{\de k_x}{2\pi m_{\rm FP}^2}\Bigg(\frac{\sin{\left(k_x(\eta-\eta')\right)-\cos{\left(\sqrt{k_x^2+m_{\rm FP}^2}(\eta'-\eta_0)\right)}\sin{\left(k_x(\eta-\eta_0)\right)}}}{k_x}\\
& \quad+\frac{\cos{\left(k(\eta-\eta_0)\right)}\sin{\left(\sqrt{k_x^2+m_{\rm FP}^2}(\eta'-\eta_0)\right)}}{\sqrt{k_x^2+m_{\rm FP}^2}}\Bigg)e^{i k_x (x-x')}\\
& =\frac{\text{sgn}(x-x'+\eta-\eta')-\text{sgn}(x-x'-\eta+\eta')}{4m_{\rm FP}^2}+\frac{i}{8\pi\, m_{\rm FP}^2}\Big[J(\Delta \eta',\nu)+J(-\Delta \eta',\nu)\\
& \quad-J(\Delta \eta',\tilde{\nu})-J(-\Delta \eta',\tilde{\nu})-I_1(\Delta \eta',\nu)+I_1(-\Delta \eta',\nu) -I_1(\Delta \eta',\tilde{\nu})+I_1(-\Delta \eta',\tilde{\nu})\Big]~,
\end{split}
\end{equation}
where sgn is the sign function with $\text{sgn(0)}=0$, $\Delta \eta'=\eta'-\eta_0>0~,\;\nu=x-x'+\eta-\eta_0~,\;\tilde{\nu}=x-x'-\eta+\eta_0$, with the auxiliary function
\begin{equation}\label{Eq: definitonIk}
    J(p,q)=\int\limits_{-\infty}^{+\infty}\text{d}k_x\,\frac{e^{i (p\,\omega_k+q\,k_x)}}{k_x}~.
\end{equation}
Therefore, to obtain an analytic expression for the correlation of $u$ and $v$, it remains only to compute the integral $J(p,q)$. The integral $I_1(p,q)$ has already been evaluated in Sec.~\ref{App:autocorrelation-v}, with its exact form given in Eq.~\eqref{Eq:I1-exact}, and its asymptotic approximation in Eq.~\eqref{Eq:I1-approx}. Again, after complex integration, one obtains
\begin{equation}
    J(p,q)+J(-p,q)=\left\{\begin{array}{ll}
         2\pi i\,\text{sgn}(q)\cos{\left(m_{\rm FP} \,p\right)}& \text{for }  |p|\leq |q|~,\\
        2i\int\limits_{-m_{\rm FP}}^{m_{\rm FP}}\text{d}u \, \dfrac{e^{-u\,q}}{u}\sin{\left(\sqrt{m_{\rm FP}^2-u^2}|p|\right)}& \text{for }  |p|>|q|~.
    \end{array}\right.
\end{equation}
For largely separated events, satisfying $m_{\rm FP}|p|\gg 1$, we have the asymptotics
\be
 J(p,q)+J(-p,q)\approx 2\pi i\, \text{sgn}(q) \cos(m_{\rm FP}\,p)-2i\sqrt{\frac{2\pi}{m_{\rm FP}}}\frac{p}{q}\left(p^2-q^2\right)^{-1/4}\sin\left(m_{\rm FP}\sqrt{p^2-q^2}+\frac{\pi}{4}\right)~.
\ee

Therefore, for the covariance of $u$ and $v$ we have: 
\begin{itemize}
    \item timelike or lightlike separated events, $|\eta-\eta'|\geq|x-x'|$:
    \begin{equation}
        \begin{split}
& \langle u(\eta,x)v(\eta',x')\rangle-\langle u(\eta,x)\rangle\langle v(\eta',x')\rangle \\
& = -\frac{2\noise^2\sin(2\theta)}{M_g^4} \frac{\text{sgn}(\Delta x+(\eta-\eta'))-\text{sgn}(\Delta x-(\eta-\eta'))-2\cos{\left(m_{\rm FP}(\eta'-\eta_0)\right)}}{4m_{\rm FP}^2}\Theta(\eta-\eta')\\
& \quad +\,(\eta\stackrel{\text{sym}}{\longleftrightarrow}\eta')~.
\end{split}
\end{equation}
To obtain this expression we had to write the inequality $|\eta-\eta'|\geq|x-x'|$ in terms of $\Delta\eta'$, $\nu$, and $\tilde{\nu}$. It can be shown that this case corresponds to either $\nu\geq|\Delta\eta'|$ and $|\tilde{\nu}|>|\Delta\eta'|$ or $\nu>|\Delta\eta'|$ and $|\tilde{\nu}|\geq|\Delta\eta'|$, with $\nu>0$ and $\tilde{\nu}<0$. Note also that this result is exact.
\item spacelike separated events with a common causal past, $|\eta-\eta'|<|x-x'|< \eta+\eta'-2\eta_0$:
\begin{equation}
    \begin{split}
&\langle u(\eta,x)v(\eta',x')\rangle-\langle u(\eta,x)\rangle\langle v(\eta',x')\rangle\\
& \approx-\frac{2\noise^2\sin(2\theta)}{M_g^4} \frac{(\eta+\eta'-2\eta_0-|\Delta x|)\sin{\left(\frac{\pi}{4}+m_{\rm FP}\sqrt{(\eta'-\eta_0)^2-(\eta-\eta_0-|\Delta x|)^2}\right)}}{2\sqrt{2\pi\,m_{\rm FP}^5}(\eta-\eta_0-|\Delta x|)[(|\Delta x|-(\eta-\eta'))(\eta+\eta'-2\eta_0-|\Delta x|)]^{1/4}}\Theta(\eta-\eta')\\
& \quad +\, (\eta\stackrel{\text{sym}}{\longleftrightarrow}\eta')~.
\end{split}
\end{equation}
In terms of the above-defined quantities, this case corresponds to $\nu>|\Delta\eta'|$ and $\tilde{\nu}<|\Delta\eta'|$ for $\nu>0$ and $\tilde{\nu}>0$, or $|\nu|<|\Delta\eta'|$ and $|\tilde{\nu}|>|\Delta\eta'|$ for $\nu<0$ and $\tilde{\nu}<0$.
\item spacelike separated events without a common causal past,$|x-x'|\geq \eta+\eta'-2\eta_0>\eta-\eta'$: 
    \begin{equation}
        \langle u(\eta,x)v(\eta',x')\rangle-\langle u(\eta,x)\rangle\langle v(\eta',x')\rangle=0~.
    \end{equation}
    This situation corresponds to $\nu>|\Delta\eta'|$ and $\tilde{\nu}\geq|\Delta\eta'|$ for $\nu>0$ and $\tilde{\nu}>0$, or $|\nu|\geq|\Delta\eta'|$ and $|\tilde{\nu}|>|\Delta\eta'|$ for $\nu<0$ and $\tilde{\nu}<0$.
\end{itemize}
Finally, to compute the correlation of $u$ and $v$ from \eqref{Eq:definition-correlation}, we will also need the variance of $u$ and $v$ obtained above, given by \eqref{Eq: variance-u} and $\eqref{Eq: variance-v}$, respectively.

\bibliographystyle{bib-style}
\bibliography{references}
\end{document}